\documentclass[useAMS,usenatbib]{mn2e}
\usepackage{graphicx}

\title[SN\,2011hs]{SN\,2011hs: a Fast and Faint Type IIb Supernova from a Supergiant Progenitor}
\author[F. Bufano et al.]{F. Bufano$^{1}$\thanks{E-mail:
milena.bufano@gmail.com},
  G. Pignata$^{1}$, M. Bersten$^{2}$, P. A. Mazzali$^{3,4,5}$, S. D. Ryder$^{6}$, R. Margutti$^{7}$, 
\newauthor D. Milisavljevic$^{7}$, L. Morelli$^{8}$, S. Benetti$^{5}$, E. Cappellaro$^{5}$, S. Gonzalez-Gaitan$^{9}$, 
\newauthor C.\ Romero-Ca\~nizales$^{10}$, M. Stritzinger$^{11}$,  E. S. Walker$^{12}$, J. P. Anderson$^{9}$, C. Contreras$^{11,13}$,  
\newauthor T. \rm{de} Jaeger$^{9}$, F. F\"orster$^{9}$, C. Gutierrez$^{9}$,  M. Hamuy$^{9}$, E. Hsiao$^{13}$, N. Morrell$^{13}$,   
\newauthor F. Olivares E.$^{1}$, E. Paillas$^{1}$, S. Parker$^{14}$, E. Pian$^{15,16}$, T. E. Pickering $^{17}$,  N. Sanders$^{7}$,   
\newauthor C. Stockdale$^{6,18}$, M. Turatto$^{5}$, S. Valenti$^{5}$, R. A. Fesen$^{19}$, J. Maza$^{9}$,  K. Nomoto$^{2}$,
\newauthor M. M. Phillips$^{13}$, A. Soderberg$^{7}$ \\
$^{1}$Departamento de Ciencias Fisicas, Universidad Andres Bello, Avda. Republica 252, Santiago, Chile\\
$^{2}$Kavli Institute for the Physics and Mathematics of the Universe (WPI), The University of Tokyo, Kashiwa, Chiba 277-8583, Japan\\
$^{3}$Astrophysics Research Institute, Liverpool John Moores University, Liverpool, UK\\
$^{4}$Max-Planck Institut f¬ur Astrophysik, Karl-Schwarzschildstr. 1, D-85748 Garching, Germany\\
$^{5}$INAF-Osservatorio Astronomico di Padova, Vicolo dell'Osservatorio 5, 35122 Padova, Italy\\
$^{6}$Australian Astronomical Observatory, P.O. Box 915, North Ryde, NSW 1670, Australia\\ 
$^{7}$Harvard-Smithsonian Center for Astrophysics, 60 Garden Street, Cambridge, MA 02138, USA\\
$^{8}$Dipartimento di Fisica e Astronomia ``G. Galilei", Universit\`a di Padova, Vicolo dell'Osservatorio 3, I-35122 Padova, Italy\\
$^{9}$Departamento de Astronomia, Universidad de Chile, Casilla 36-D, Santiago, Chile\\
$^{10}$Instituto de Astrof\'{\i}sica, Facultad de F\'{\i}sica, Pontificia Universidad Cat\'olica de Chile, Casilla 306, Santiago 22, Chile \\
$^{11}$Department of Physics and Astronomy, Aarhus University, Ny Munkegade, DK-8000 Aarhus C, Denmark\\
$^{12}$Yale University, Department of Physics, P.O. Box 208120, New Haven, CT 06520-8120, USA\\
$^{13}$Carnegie Observatories, Las Campanas Observatory, Colina El Pino, Casilla 601, Chile\\
$^{14}$Backyard Observatory Supernova Search, Oxford, Canterbury, New Zealand\\
$^{15}$Scuola Normale Superiore di Pisa, Piazza dei Cavalieri 7, I-56126 Pisa, Italy\\
$^{16}$INAF-Istituto di Astrofisica Spaziale e Fisica Cosmica, Via P. Gobetti 101, I-40129 Bologna, Italy\\
$^{17}$Southern African Astronomical Observatory, Observatory Road, Observatory 7925, South Africa\\
$^{18}$Physics Dept., Marquette University, P.O. Box 1881, Milwaukee, WI~53201, U.S.A.\\
$^{19}$Department of Physics and Astronomy, Dartmouth College, 6127 Wilder Lab, Hanover, NH 03755, USA
}

\begin{document}

\date{Accepted ... Received ..; in original form ...}

\pagerange{\pageref{firstpage}--\pageref{lastpage}} \pubyear{2013}

\maketitle

\label{firstpage}

\begin{abstract}
Observations spanning a large wavelength range, from X-ray to radio, of the Type IIb supernova 2011hs are presented, covering its evolution during the first year after explosion. The optical light curve presents a narrower shape and a fainter luminosity at peak than previously observed for Type IIb SNe. 
High expansion velocities are measured from the broad absorption H\,I and He\,I lines.  
 From the comparison of the bolometric light curve and the time evolution of the photospheric velocities with hydrodynamical models, we found that SN\,2011hs is
 consistent with the explosion of a 3--4\,M$_\odot$ He-core progenitor star, corresponding to a main sequence mass of 12--15\,M$_\odot$,
 that ejected a  mass of $^{56}$Ni of about 0.04\,M\,$_\odot$, with an  energy of $E= 8.5 \times 10 ^{50}$ erg. 
 Such a low-mass progenitor scenario is in full agreement with the modelling of the nebular spectrum taken at $\sim$215 days from maximum.
  From the modelling of the adiabatic cooling phase, we infer a progenitor radius of $\approx$500--600 $R_\odot$, clearly pointing to an extended progenitor star.
The radio light curve of SN\,2011hs yields a peak luminosity similar to that of SN\,1993J,
but with a higher mass loss rate and a wind density possibly  more similar to that of SN\,2001ig. 
Although no significant deviations from a smooth decline have been found in the radio light curves, 
we cannot rule out the presence of a binary companion star.
\end{abstract}


\section{Introduction}
Core collapse supernovae (CC SNe) represent the final stage of the evolution of zero main sequence (ZAMS) massive stars  $\ga 8 M_\odot$ \citep{Heger03}. SNe are generally classified  on the basis of their early spectral appearance, which gives an indication on the nature of the evolutionary phase of the progenitor stars at the time of their explosion.
The main division is defined by the presence or absence of hydrogen (H) lines, which splits CC SNe in Type II and Type Ib/c, respectively, and reveals whether or not the progenitor retained its H envelope before the explosion. 
An interesting group of CC SNe undergoes a peculiar spectral metamorphosis during their evolution: their spectra  present at early phases broad H\,I absorption lines like a Type II SNe, which later disappear while He\,I features become predominant like in stripped envelope  (SE) SN Type Ib spectra. For this reason, they are called Type IIb SNe.\\
 The mechanism that explains how the progenitors of Type IIb SNe could shed most of the H layer at the time of explosion, while retaining enough mass ($\la$1 M$_\odot$; \citealt{Nomoto93J}) to show H signatures in their spectra is still under debate. 
The  proposed scenario points to the explosion of a relatively high mass star  ($\sim$ 25--30 M$_\odot$), which lost its H envelope by radiatively driven winds (\citealt{kw07}; \citealt{sto07}; \citealt{SmithWR}), or, alternatively, by mass transfer to a binary companion star \citep{Yoon10}.
A close binary companion could strip most of the external envelope also of a less massive  star,  allowing  stars
with a larger radius (like supergiant stars) to explode as Type IIb SNe (\citealt{Eldridge}; \citealt{SmithCCfrac}; \citealt{Benvenuto}). 
In the extreme case of such a mass transfer, the companion could even
spiral into the primary star and remove a large fraction of the envelope to form a single star progenitor of a Type IIb SN  \citep{Nomoto95}.
Recently, it has been argued that a single star with an initial mass of 12--15 M$_\odot$
could explode as a  Type IIb SN,  if a much higher mass loss wind (up to 10 times) than the standard one (\citealt{DeJager}; \citealt{Mauron}) is assumed, but the possible physical  mechanism powering such a strong wind is still unidentified  (\citealt{Georgy12}).
So far, the wide variety in the observational properties of the small number of well-observed Type IIb SNe has made it impossible to  favor a particular scenario among the possible ones.\\
SN 1987K was the first SN showing the Type II-Ib transition \citep{Filippenko},  although the most known and best studied Type IIb is SN 1993J (\citealt{Filippenko93J}, \citealt{Barbon93J}, \citealt{Richmond93J}), considered the prototype of this class.  
The progenitor star of SN\,1993J has been identified  in pre-explosion images as a K-type supergiant in a binary system (\citealt{Aldering93J}, \citealt{Maund93J}). Interestingly, a yellow supergiant (YSG) star was identified in  pre-explosion images as the progenitor star of another well-studied Type IIb, SN\,2011dh (\citealt{Maund11dh}; \citealt{VanDyk11dh}; \citealt{Arcavi11dh}; \citealt{Howie11dh}; \citealt{Sahu11dh}; \citealt{Ergon11dh}),  fully consistent with the numerical modelling of its bolometric light curve  (\citealt{Bersten}),  and definitely confirmed  by the disappearance of the YSG candidate in post-explosion images taken almost two years after its explosion (\citealt{Ergon11dh}; \citealt{VanDyk11dhprog}). 
Evidence of a binary companion has also been claimed for the Type IIb SN\,2001ig (\citealt{ryd04}, 2006).\\
On the other hand,  for the Type IIb SN\,2008ax pre-explosion colours favour  a bright stripped-envelope massive star with initial mass between 20--25\,M$_\odot$,
although  the  possibility of an interacting binary in a low-mass cluster could not be ruled out (\citealt{Crockett08ax}; \citealt{Chornock08ax}; \citealt{Pasto08ax}; \citealt{Taubix08ax}). \\
Extensive data sets have been published for only a handful of Type IIb SNe,  mainly due to the limited number of Type IIb SNe discoveries. 
This is the consequence of the intrinsic low rate (Type IIb SNe represent just 12\% of the observed CC SNe; \citealt{Li_rate})
and of possible misclassifications as Type Ib SNe, due to the strong  dependence of the H lines strength  on its mass/distribution and on the phase at which the SN is discovered (see \citealt{Chornock08ax}; \citealt{Max07Y}; \citealt{Dan11ei}). \\
In this paper, we present the results obtained from the analysis of the data collected during the multi-wavelength followup campaign of the Type IIb SN\,2011hs, located at  R.A.=22$^h$57$^m$11$\fs{77}$ and Decl.=$-$43$^\circ$23$'$04$\farcs$08  (equinox
2000.0) at 20$''$ west and 41$'' $north of the nucleus of the galaxy IC\,5267.  
SN\,2011hs was discovered on Nov. 12.5 (UT dates are used throughout the paper)
with a  35-cm Celestron C14 reflector (+ ST10 CCD camera), at an unfiltered magnitude of 15.5 \citep{CBET}.
Our observational campaign started immediately after the announcement of the discovery, 
sampling the SN evolution in a wide wavelength range, spanning from the X-ray to the radio domain. 
The first optical spectrum obtained  on Nov. 14.9  with the 10-m SALT telescope (+RSS), revealed that SN\,2011hs was
a Type IIb SN, with a H$\alpha$ expansion velocity resembling the fast expanding Type IIb SN\,2003bg \citep{CBET}.\\
X-ray and ultraviolet (UV) observations of SN\,2011hs were secured with the {\it Swift}  \citep{Gehrels04} satellite from 
Nov. 15 until the SN faded below the detection threshold.\\
We intensively monitored  the optical and near-infrared (NIR) spectrophotometric evolution of SN\,2011hs  out to
 $\sim65$\,days  past the discovery, when the campaign was suspended because of the SN conjunction with the Sun, and
  then restarted in April 2012  and continued until 2012 Oct. 22, the epoch of the last observation published here.
The presented optical/NIR dataset is the outcome of the coordination of various  observing programs at 
different telescopes in different observatories located in Chile [Las Campanas Observatory (LCO), ESO La Silla Observatory and Cerro Tololo Interamerican Observatory (CTIO)] 
and South Africa (Southern African Astronomical Observatory). \\
Radio monitoring of SN\, 2011hs with the Australia Telescope Compact
Array\footnote{The Australia Telescope is funded by the Commonwealth
  of Australia for operation as a National Facility managed by CSIRO.}
(ATCA) began within a week from its discovery, collecting multi-frequency radio flux data for
the first 6\, months at frequencies between 1 and 20\, GHz. \\
The paper is organized as follows: the photometric and spectroscopic observations are presented in Section \ref{obsevations_sect}, where the observational campaign and methods for data reduction for each wavelength range are described. In Section \ref{galaxy}, we  define  the host galaxy properties, i.e. distance and dust extinction, while
in  Section \ref{lc_sec} and Section \ref{SpectralEvolution} we analyze the photometric and spectroscopic evolution of the SN. In Section \ref{bolom_sec}, we present
the hydrodynamical  modelling performed to estimate physical parameters of the SN progenitor and its explosion.
Section \ref{radio} deals with the radio data modelling and, finally, in Section \ref{summary} we summarize the results and present our conclusions.


\begin{figure}
\begin{center}
\includegraphics [width=0.5\textwidth]{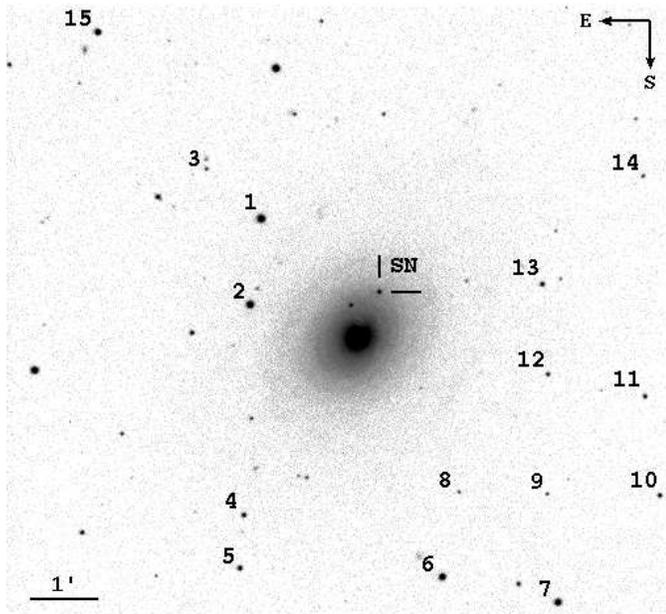}
\caption{SN 2011hs in IC 5267 and the sequence of local reference stars (cf. Table\,\ref{localJohn} and \ref{localSloan}). The image is a $V$ frame taken with
PROMPT on 2011 Nov 19.}
\label{FC}
\end{center}
\end{figure}


\section{Observations}\label{obsevations_sect}
\subsection{ X-Ray Observations}
{\it Swift}-XRT \citep{Burrows05} observations were acquired starting from 
Nov. 15.3 to Nov. 26.3, for a total of 31.3 ks.
HEASOFT (v. 6.12) package has been used to calibrate and analyze {\it Swift}-XRT  data. 
Standard filtering and screening criteria have been
applied. We find evidence for X-ray emission originating from the host
galaxy  nucleus at the level of $\sim(2.2\pm0.2)\times
10^{-13}\,\rm{erg\,s^{-1}cm^{-2}}$ (unabsorbed flux in the 0.3-10 keV
energy band). As reported in \citet{MarguttiAtel}, no significant X-ray emission is detected at the SN
position, with a 3 sigma upper limit of $1.3\times10^{-3}\,\rm{c\,s^{-1}}$ 
(0.3-10 keV band).
 The Galactic neutral hydrogen column density in the direction 
of the SN is $1.22\times10^{20}\,\rm{cm^{-2}}$ \citep{Kalberna05}.
Assuming a spectral photon index $\Gamma=2$, this translates into an unabsorbed upper limit flux
$F<6.4\times10^{-14}\,\rm{erg\,s^{-1}cm^{-2}}$, corresponding to a luminosity 
 $L<5.2\times10^{39}\,\rm{erg\,s^{-1}}$ at the assumed distance of 
 26.4 Mpc (see Sect.\,\ref{galaxy}).  The presence of an extended X-ray emission from the host
galaxy nucleus and the proximity of SN\,2011hs to the nucleus (compared with the
{\it Swift}-XRT PSF) prevents us from providing a deeper
limit. In any case, considering the X-ray luminosities observed in previous studied Type IIb SNe,
we can conclude that a SN like SN\,2011dh, with a $L_{X[0.3-8keV]}\approx1.5\times10^{38}\,\rm{erg\,s^{-1}}$ \citep{Soderberg12}
would not have been detected, whilst SN\,1993J, with a luminosity of $L_{X[0.3-8keV]}\approx8\times10^{39}\,\rm{erg\,s^{-1}}$,
would stand out on the background.

\subsection{ UVOT Observations}
{\it Swift}-UVOT \citep{Roming05} data were acquired using the 
6 broad-band filters ({\it w2}, {\it w1},{\it m2}, {\it u}, {\it b} and {\it v}), spanning a wavelength range
from $\lambda_c=1928$ \AA\ ({\it w2} filter) to $\lambda_c=5468$ \AA\ ({\it v} filter). 
Data have been analyzed following 
the prescriptions of \cite{Brown09}. In particular, a 3$''$ aperture 
has been used to maximize the signal-to-noise ratio and limit the 
contamination from host galaxy light.  We removed the residual
contamination from host galaxy light 
selecting a number of background regions close to the SN site
 (unfortunately, no UVOT pre-explosion images of SN\,2011hs are available).
{\it Swift} UV and optical photometry, based on the UVOT photometric system of \cite{Poole08}, is reported in Tabs. \ref{UV_mag} and \ref{optical_mag}.

\subsection{Optical and NIR Photometry}\label{photometry}
The optical photometric follow up of SN\,2011hs was almost entirely obtained with the 0.41m Panchromatic Robotic Optical Monitoring and Polarimetry Telescope (PROMPT; \citealt{Reichart})  and the 50cm CATA500 Telescope, both located at the CTIO, which offered {\it BVRI}+{\it u$^\prime$g$^\prime$r$^\prime$i$^\prime$z$^\prime$} and {\it BV}+{\it u$^\prime$g$^\prime$r$^\prime$i$^\prime$}  coverage, respectively. Additional optical ({\it UBVRI}) data were acquired at the du Pont Telescope, a 2.5m-class telescope at LCO, using the Wide Field Reimaging CCD Camera (WFCCD),  and at the 3.6m ESO New Technology Telescope (NTT), mounted with the ESO Faint Object Spectrograph and Camera 2 (EFOSC2). The last nebular $BVRI$ epoch was
acquired at the ESO Paranal Observatory with the Very Large Telescope Unit 1 (VLT/UT1) equipped with the FOcal Reducer/low dispersion Spectrograph 2 (FORS2).  
For optical images, standard reductions were performed using IRAF\footnote{IRAF is distributed by the National Optical Astronomy Observatories, which are operated by the Association of Universities for Research in Astronomy, Inc. under contract with the National Science Foundation.} tasks, including bias and flat-field corrections. A point-spread-function (PSF) fitting method was applied to measure the SN magnitudes.
Since the background at the SN position is fairly regular, we did not need to perform a template subtraction, and thus we removed its contribution to the SN flux evaluating it by means of a  two-dimensional low-order polynomial fit of the region surrounding the SN.
SN\,2011hs instrumental magnitudes were calibrated to the standard Johnson--Cousins and Sloan photometric systems using the relative colour equations,
obtained for each instrument  observing \citet{Landolt07} and \citet{Sloan} standard star fields over the course of photometric nights.  
We also measured and calibrated the magnitudes of a local sequence of stars, whose positions in the field are shown in Fig.\ref{FC}.  Magnitudes were computed via a weighted average of the measurements made during the photometric nights (3 nights for $UBg^\prime$ bands and 6 nights for the remaining ones) and reported in Tables\,\ref{localJohn} and \ref{localSloan}. We used the magnitudes of the  local  stars sequence to obtain the photometric zero-points for non-photometric nights. 
Because of the small field of view and of the faintness of the local field stars, only one star was available for the calibration of the U-filter images. \\
During the very early phases unfiltered images were collected by amateur astronomers of the Backyard Observatory Supernova Search (BOSS\footnote{http://www.bosssupernova.com/}) project. We included  
BOSS observations performed  on Nov. 12.5 (used for the discovery; \citealt{CBET}) and Nov 14.15 and 14.16 in our analysis.
Since the quantum efficiency of the employed CCD peaks around 6500\,\AA, we calibrated the unfiltered magnitudes as  Johnson-Bessell $R$ band images.
In addition, by using the local stars sequence we estimated a color correction, which turned out to be quite small ($<$20\% of the ($V-R$) colour) and
corresponding to a negligible correction for the  SN $R$ band magnitudes ($\sim0.02-0.03$ mag).\\ 
NIR photometry was acquired, using a variety of facilities, including  {\it JHK} images with the SOFI camera mounted at NTT; {\it JH} images with the NIR camera of the robotic 60-cm telescope REM in LaSilla; and finally, {\it JH} images with the RetroCam at the DuPont telescope.  
The images from each instrument were calibrated using field stars from  the Two Micron All-Sky Survey (2MASS). 
SN\,2011hs magnitudes are tabulated in Tables\,\ref{optical_mag}, \ref{sloan_mag} and \ref{NIR_mag}, where the uncertainties are a quadratic sum of the errors of the instrumental SN magnitude measurement and the photometric calibration.

\subsection{Optical and NIR Spectroscopy}
The log of the 24 spectroscopic epochs we obtained is reported in Table \ref{spec_journal}.
In addition to the telescope/instruments used also for the photometry (and described in Sect. \ref{photometry}), optical spectra were taken at: the  
 Southern African Large Telescope (SALT) with the Robert Stobie Spectrograph (RSS); the Magellan Telescope (+ LDSS3) at LCO; and  the SOAR(+Goodman Spectrograph) Telescope  on Cerro Pachon. 
 NIR spectra were obtained at NTT+SOFI and at the Magellan Telescope + FIRE spectrograph.\\
 Spectra were reduced using IRAF tasks, within noao.onedspec and ctioslit package. Spectrophotometric (\citealt{Hamuy92}, 1994) and telluric standard-star exposures taken on the same night as the SN 2011hs observations were used to flux-calibrate the extracted spectra and to remove telluric absorption features, respectively. We checked
 the flux calibration of the spectra against the simultaneous broad-band photometry and, if required, the spectrum was rigidly scaled to match 
 the photometry. 
The simultaneous spectra obtained at Magellan(+IMACS) and at the du Pont Telescope (+WFCCD) on Nov. 18th (see Table \ref{spec_journal}) were
combined and presented as a single spectrum.

 \subsection{Radio Observations}\label{s:radio}
SN\,2011hs follow-up at the radio wavelength range (2.0-18.0 GHz) was performed at the Australia Telescope Compact Array (ATCA) and started from  Nov 17.4 UT.
The total time on-source ranged from 1 to 3 hours, yielding sufficient $uv$-coverage to comfortably resolve SN\, 2011hs
from the nucleus of IC\, 5267  ($45\arcsec$ to the south-east). 
Table\,\ref{t:radiofluxes} contains the complete log of observations and radio flux measurements from the ATCA,
where epochs are given as days elapsed since the discovery. 
The Compact Array Broad-band Backend (CABB; \citealt{cabb11}) provides $2
\times 2$\, GHz IF bands, each of which has $2 \times 2048$ channels of
1\, MHz each. The first 2 epochs covered the frequency bands of
4.5--6.5\, GHz and 8.0--10.0\, GHz; the next 5 epochs also included the
bands 16.0-18.0\, GHz and 18.0--20.0\, GHz; while the final 3 epochs
replaced these highest frequency bands (where the SN was no longer
detectable) with both IFs now covering 1.1--3.1\, GHz.\\
The ATCA primary flux calibrator, PKS\, B1934-638 was observed once
per run at each frequency to set the absolute flux scale.
It also defined the bandpass calibration in each band, except for
18\, GHz where the brighter source PKS\, B1921-293 was used instead.
Frequent observations of the nearby source PKS\, B2311-452 allowed us to
monitor and correct for variations in gain and phase during each run,
and to update the antenna pointing model at 18\, GHz.

The data for each observation and separate IF band have been edited
and calibrated using tasks in the {\sc miriad} software package
\citep{mir95}. The large fractional bandwidths used enable
``multi-frequency synthesis'', in which $uv$-plane coverage is
improved by gridding each channel individually, followed by
multi-frequency deconvolution of the dirty image to account for the
spectral index of each source. Despite this the factor of 3 change in beam
size over the 1.1--3.1~GHz band, coupled with the significant amount
of interference wiping out the lower 512~MHz of this band,
required its splitting into two sub-bands of 768~MHz, centered on
2.0~GHz and 2.7~GHz.

Robust weighting was employed in the imaging to give the best
compromise between the minimal sidelobes produced by uniform
weighting, and the minimal noise achieved with natural weighting.
While Gaussian fitting of the clean beam to an unresolved source is a
standard way of determining the flux of a radio point source, at low
flux levels fitting to the calibrated visibilities in the $uv$-dataset
can be more reliable. The UVFIT task has been used to fit simultaneously a point source at
the known location of SN~2011hs, as well as the nearby nucleus of the
host galaxy IC~5267, which was of comparable but more stable
luminosity than SN~2011hs.  Following \citet{kw11} the uncertainties in
Table~\ref{t:radiofluxes} are the quadrature sum of the image rms and
a fractional error on the absolute flux scale in each band.

\section{Distance and dust extinction}\label{galaxy}

The adopted distance for SN\,2011hs is based on the  recession  velocity value
obtained from the accurate folding of the stellar velocity rotation curve \citep{Morelli08}. 
We derived a recession heliocentric velocity of 1710 $\pm$ 20 \,km\,s$^{-1}$,  in excellent agreement with the tabulated value in the NED catalog \citep{NED}.  Corrected for the peculiar solar motion \citep{kerr},  this corresponds to a redshift equal to z = 0.0057 $\pm$ 0.0001
and a distance modulus $\mu=31.85\pm 0.15$ mag (where Hubble constant H$_0$= 73 \, km\,s$^{-1}$\,Mpc$^{-1}$,  $\Omega_{\Lambda}$=0.73 and $\Omega_{M}$=0.27).  
On the other hand, from  the early time SN\,2011hs spectra, we measured an average shift of the H$\alpha$ central wavelength 
 corresponding to a recession heliocentric velocity of 1910 $\pm$ 40\,km\,s$^{-1}$, 200\,km\,s$^{-1}$ higher than the galaxy nucleus.
This is consistent with the values obtained  from the velocity rotation curve of the gas ([N II] line) component at a radius of 45\,arcsec along 
the major axis of the galaxy.  The data reduction and analysis used to extract the gas kinematic are described in \citet{Morelli08} and \citet{Morelli12}. The gas and stellar kinematic for this S0 galaxy unveiled a very peculiar and interesting
structure, with the inner (r$<$ 5\,arcsec) region of the galaxy rotating in the opposite direction with respect to its external region
(as seen in e.g. NGC 4826, \citealt{Rubin}). 
The kinematics of the stellar and gas components show the same radial trend, although the values of the stellar and gas velocity do not match
in the disk dominated region, being $<$30\,km\,s$^{-1}$ and $>$100\,km\,s$^{-1}$, respectively. 
This suggests that the face on stellar component is not aligned with the gas components, 
probably due to a strong warp in the radial structure of the gas disk or to a different inclination of stellar and gas disks.
It is not obvious how to associate the SN to either the stellar or the gaseus component
and decide which velocity correction to apply to our spectra.
Nevertheless,  considering that the $\Delta$v$\sim$200\,km\,s$^{-1}$ with respect to the nucleus value would result in a wavelength offset 
of $\sim$4\,\AA\ (which is much greater than the $\sim$1--2\,\AA\  uncertainty affecting the spectral wavelength calibration), we found that by applying  the shift found from the H$\alpha$ emission, the maxima of the forbidden emission lines  in the nebular spectra (see Sect.\,\ref{nebular_sec}) fall at the proper rest-frame wavelengths. Thus  a recession  velocity of $v$=1910 $\pm$ 40\,km\,s$^{-1}$ will be employed for the analysis in this paper. \\
For the Milky Way  extinction we used  the IR maps by \citet{Schlafly}, that give for our observing direction an $E(B-V)_{\rm MW} = 0.011 \pm 0.002$ mag.
On  the other hand, the estimation of the  extinction  due to the host galaxy dust  is a tricky issue, and is a considerable source of 
 systematic uncertainty in SN studies. 
A proxy for host galaxy reddening is given by a correlation that links the column density of neutral sodium (Na) with
the absorption and scattering properties of dust, using the equivalent width (EW) of the interstellar Na\,I\,D absorption doublet 
($\lambda\lambda$5890, 5896 \AA; \citealt{TurattoNaI}; \citealt{Poznanski_rel}). 
We estimated  EW(Na\,I\,D)  from the first spectrum ($-$5 days) to 10 days after the $B$-maximum light (hereafter $t(B)_{max}$), when
He\,I lines started to dominate the spectrum (cf. Sec. \ref{SpectralEvolution}).
We observed a high scatter of the EW(Na\,I\,D) measurements, although it did not show any  trend. Thus we adopt
an average value EW(Na\,I\,D)= 0.90$\pm$0.19\,\AA\ (with the error given by the rms of the distribution).
Applying the relation by \citet{Poznanski_rel}, $log(E(B-V))= 1.17 \times {\rm EW(Na\,I\,D)}-1.85\pm0.08$,
we obtain an  $E(B-V)_{\rm host} = 0.16 \pm 0.07$ mag.
As a comparison, applying the relation by \citet{TurattoNaI},  ${\rm E(B-V)=0.16\times EW(Na\,I\,D)}$,
an E(B-V)=0.14$\pm$0.03 mag is obtained, in agreement with the previous value.
Thus the total (Milky Way + host galaxy) color excess value adopted throughout this work is 
$E(B-V)_{tot}=0.17 \pm 0.08$ mag. 


\begin{figure}
\begin{center}
\includegraphics [width=0.5\textwidth]{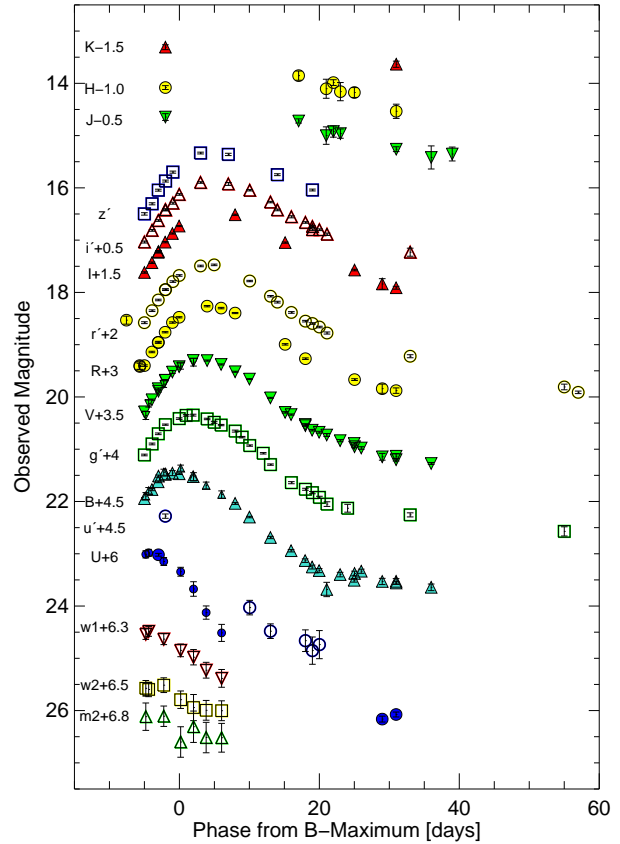}
\caption{Early-phase light curves of SN\,2011hs extending from the UV to the NIR wavelengths. Observed magnitudes are not corrected for reddening and vertically shifted for clarity of the plot. The epochs are computed with respect to the $t(B)_{max}$, i.e. Nov. 20 (2,455,885.5 $\pm$ 1.0 JD). The filter name corresponding to each light curve is reported on the left side. {\it Swift}/UVOT {\it u, b} and {\it v} magnitudes are reported with small symbols. }
\label{lightcurve}
\end{center}
\end{figure}



\begin{figure}
\begin{center}
\includegraphics [width=0.5\textwidth]{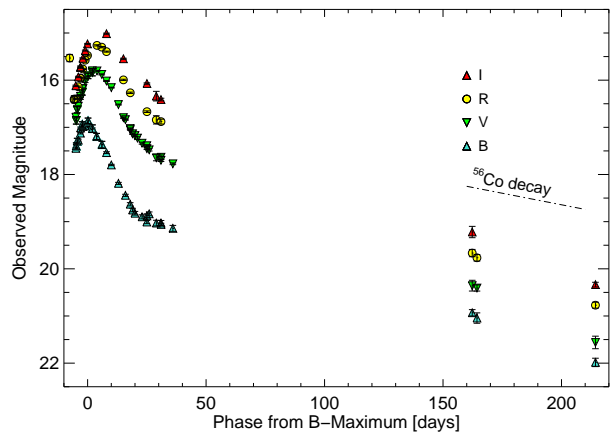}
\caption{Time evolution of SN\,2011hs in {\it BVRI} bands through the  nebular phases.}
\label{lightcurvelate}
\end{center}
\end{figure}



\begin{figure*}
\begin{center}
\includegraphics [width=10cm,angle=90]{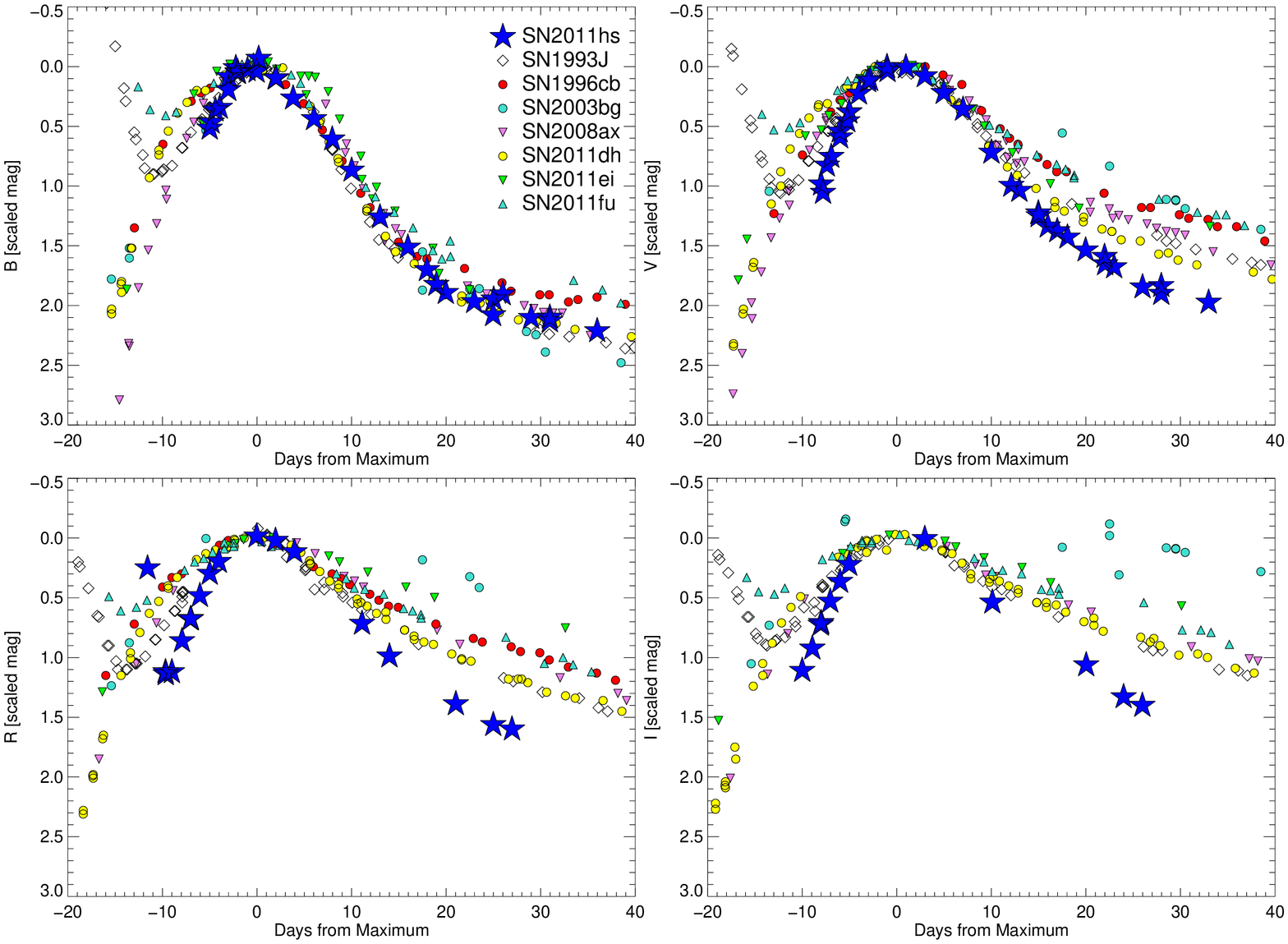}\\
\vspace{.4cm}
\caption{Comparison of SN 2011hs {\it BVRI} light curve shapes with those of SNe 1993J (\citealt{Barbon93J}; \citealt{Richmond93J}); 1996cb \citep{Qiu}; 2003bg \citep{Hamuy03bg}; 2008ax (\citealt{Pasto08ax}; \citealt{Taubix08ax}); 2011dh \citep{Ergon11dh}; 2011ei \citep{Dan11ei}; 2011fu \citep{Kumar11fu}.  The phase is computed with respect to the relative maximum light.}
\label{comp_lightcurve}
\end{center}
\end{figure*}


\section{the Light curves}\label{lc_sec}
\begin{figure}
\begin{center}
\includegraphics [width=0.49\textwidth]{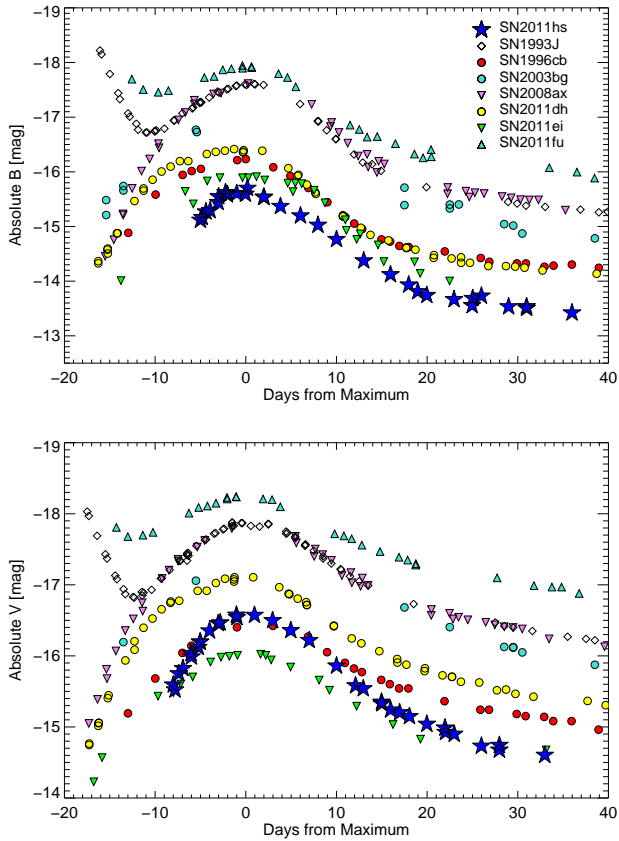}
\caption{Comparison of the $B$ ({\it upper panel}) and $V$ band ({\it lower panel}) absolute magnitude light curves of SN 2011hs compared
with those of the sample of SNe Type IIb presented in Figure\,\ref{comp_lightcurve}.}
\label{sn_lightcurve}
\end{center}
\end{figure}

Light curves of SN\,2011hs obtained in the different bands from UV to the NIR wavelength range are shown in Figs.\,\ref{lightcurve} and \ref{lightcurvelate}. The follow up spans epochs from 2011 Nov. 14.5 to 2012 Jan. 16.1 for most of the filters, with the exception of  {\it BVRI} bands for which it was extended to 2012 Jun. 20.4 UT.
As described also in Section\,\ref{bolom_sec},  Type IIb SN light curves are expected to be characterized by an initial decline,
as a consequence of the adiabatic cooling of the ejecta after the shock breakout (see e.g.  \citealt{WoosleyIIb}; \citealt{BlinnikovIIb}; \citealt{Bersten}).
Such a cooling phase has been observed in very few Type IIb SN cases: i.e. SNe 1993J (\citealt{Barbon93J}; \citealt{Richmond93J}), 2011dh \citep{Arcavi11dh} and, recently, 2011fu \citep{Kumar11fu}.
The cooling branch is followed by a rising to a peak, powered by the radioactive decay of the $^{56}$Ni,
produced during the explosion, and its daughter $^{56}$Co. 
Assuming that SN\,2011hs followed the same path,  it appears
that the observational campaign started when the SN was already
in the post-cooling rising phase (Fig.\,\ref{lightcurve}). Indeed early photometry based on the
amateur discovery image shows that  SN 2011hs initially decreases by 0.88\,mag ($R$ band) in  1.9 days.  
Such a first point, as shown in Sect.\,\ref{bolom_sec},  aids us in constraining the light curve modelling and, thus, determining the progenitor star radius at the time of the explosion. \\
 For the bands with a more detailed sampled light curve, we estimated the epoch of the peaks and their magnitude using low order polynomial fits. These are reported in  Table \ref{maximum}, along with the pre- and post-maximum light curve slope estimations
 obtained using least-squares fits.\\
SN\,2011hs reached $t(B)_{max}$ on 2011 Nov. 20 (corresponding to 2,455,885.5 JD), while light curves in redder filters peak  some days later.
As shown in Fig.\,\ref{comp_lightcurve}, SN 2011hs has the steepest rise to
 the peak among the Type IIb SN sample we found in literature (i.e. SNe 1993J, \citealt{Barbon93J}, \citealt{Richmond93J}; 1996cb, \citealt{Qiu}; 2003bg, \citealt{Hamuy03bg}; 2008ax, \citealt{Pasto08ax}, \citealt{Taubix08ax}; 2011dh, \citealt{Ergon11dh}; 2011ei, \citealt{Dan11ei}; 2011fu, \citealt{Kumar11fu}). \\
 This is true also for its post-maximum decline in the $VRI$ bands (e.g. for the I band  SN \,2011hs has a rate of  0.07$\pm$0.01 mag\,d$^{-1}$, while it is 0.05$\pm$0.01 mag\,d$^{-1}$  and 0.02$\pm$0.01 mag\,d$^{-1}$ for SN\,1993J  and SN\,2011fu, respectively; see Table 5 in \citealt{Kumar11fu}), while in the $B$ band (0.09$\pm$0.01 mag\,d$^{-1}$) it is similar to that of the other Type IIb SNe (e.g. 0.11$\pm$0.01 mag\,d$^{-1}$  and 0.10$\pm$0.01 mag\,d$^{-1}$ for SN\,1993J  and SN\,2011fu, respectively, from Table 5 in \citealt{Kumar11fu}). \\
Adopting the distance modulus and reddening discussed in Section \ref{galaxy}, we find a B and V absolute peak magnitude of $-15.63$ mag and $-16.59$ mag, respectively. A comparison among a sample of Type IIb SN absolute light curves is shown  in Fig. \,\ref{sn_lightcurve}. 
SN\,2011hs appers to be  the faintest Type IIb SN in the B band, while in the $V$ band turns out to be as bright as SN 1996cb. This results in a high $(B-V)$ colour value as shown in the next section (Sect.\,\ref{color_sec}).\\
SN 2011hs nebular photometry was  obtained only in the {\it BVRI} bands, as shown in  Fig. \ref{lightcurvelate}. 
 From it, we measured a slope of  $2.31\pm0.28$ mag\,(100d)$^{-1}$ in the $V$ band,   which is steeper
than the rate expected for $^{56}$Co decay in the case of complete $\gamma$-ray trapping (0.98 mag/100d).
This is commonly found in SE SNe at nebular phase (100-300 days after maximum) and
it is attributed to  rather low ejecta masses with  respect  to SNe IIP  \citep{Clocchiatti}.


\begin{figure}
\begin{center}
\includegraphics [width=0.45\textwidth,angle=90]{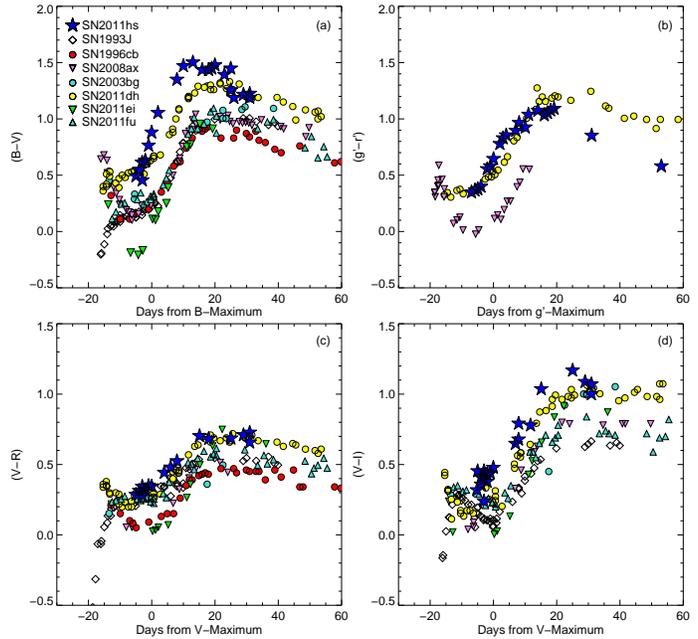}
\vspace{.4cm}
\caption{Comparison of SN 2011hs colour curves with those of the sample of SNe Type IIb presented in Figure\,\ref{comp_lightcurve}.
All the colour curves are corrected for the relative reddening value.}
\label{comp_colorcurve}
\end{center}
\end{figure}


\subsection{Colour Evolution}\label{color_sec}
In Fig. \ref{comp_colorcurve}, we show  the evolution of ($B-V$), ($g^\prime$-$r^\prime$), ($V-R$) and ($V-I$) intrinsic colour of SN\,2011hs (colour excess used $E(B-V)=0.17$ mag; see Sect.\,\ref{galaxy}) and compare them with those of the same Type IIb SNe sample, corrected for the relative reddening value.
In general, SN\,2011hs shows redder colours and more rapid evolution than the comparison SNe.
In particular, the  ($B-V$) colour evolution of most of the Type IIb SNe follow that of SN 2008ax,  that after an 
initial move to blue colour, starts getting redder with an initial rise of $\sim$0.03 mag\,d$^{-1}$  from -5 days to 5 
days,  which then  becomes steeper (with a rate of 0.056$\pm$0.004 mag\,d$^{-1}$) until about two weeks after maximum.
Only few SNe differ from this behaviour: i.e. SNe 1993J and 2011fu, which start with bluer colours at very early phases post-explosion,   
evolve monotonically to red colours, but already after $-$10 days from maximum follow the average trend,    
and SN 2011ei, whose initial rise to red colours has a steeper slope.   
 The SN 2011hs ($B-V$) colour curve (which at maximum light has an offset $\Delta(B-V)\sim$0.6 mag from SN 2008ax) displays a peculiar evolution: 
it shows an initial increase with a rate  of 0.061$\pm$0.003 mag\,d$^{-1}$ between -5 days and 15 days,
similar only to SN\,2011ei ($\Delta(B-V)/\Delta$t = 0.061$\pm$0.004 mag\,d$^{-1}$ in the same time interval);
later, during the decline to bluer color, at about one month from $t(B)_{max}$, the colour undergoes a drop of $\sim$0.2 mag in one day.
Following this, the SN 2011hs colours become similar to those of the other Type IIb SNe. 
In contrast, the $(V-R)$ and $(V-I)$ colour evolution of SN 2011hs
do not differ significantly from those of the other Type IIb SNe.
Similarly to SN 2011dh, SN 2011hs ($g^\prime-r^\prime$) colour is characterized at maximum light  by a  $\Delta(g^\prime-r^\prime)\sim$0.5 mag with respect to SN 2008ax, but it has  a much steeper slope to bluer colours during the late stages of its evolution.\\
The redder colours showed by SN 2011hs could point to an underestimation of the reddening effect. 
 It is common practice in SN studies to estimate the host galaxy extinction through the colour excess measured
by comparing the colour curve of the SN to a ``template'' curve, which is basically obtained by averaging the colour curves
of SNe belonging to the same SN class.  This method is based on two assumptions:
firstly that there is a similarity among the intrinsic colour evolutions of SNe of the same class and, secondly, that those SNe used to construct the
template curve are affected by a negligible reddening. SN\,2003bg is the only object in the well studied Type IIb SN sample that
is assumed to have a no significant extinction from the host galaxy, since its spectra do not show Na\,I\,D absorption lines \citep{Hamuy03bg}.
Unfortunately, we lack of a photometric coverage from about the maximum light epoch to 20 days later. This
prevents us to use it as a template, since we cannot ensure that its colour curve shape was similar to that of SN\,2011hs. 
Nevertheless,  we found a good agreement between the two SNe at the common epochs,
which makes us confident of the reddening correction adopted.
As we reported here, SN 2011ei had a colour evolution similar to SN 2011hs, so it could be considered a good template.
However,  it shows to be much bluer that the other Type IIb SNe  (see Fig.\,\ref{comp_colorcurve})
and, most importantly,  its reddening was estimated through the Na\,I\,D method too.
Therefore, by using it, we actually could introduce  an additional source of uncertainty in our reddening estimation. 
Thus this method cannot be applied in order to  disentangle the possible effects of the reddening from those of a different intrinsic SN colour (due to e.g. a different SN temperature) at least until a sample of unreddened Type IIb SNe will be assembled.


\begin{figure*}
\begin{center}
\includegraphics [width=0.8\textwidth]{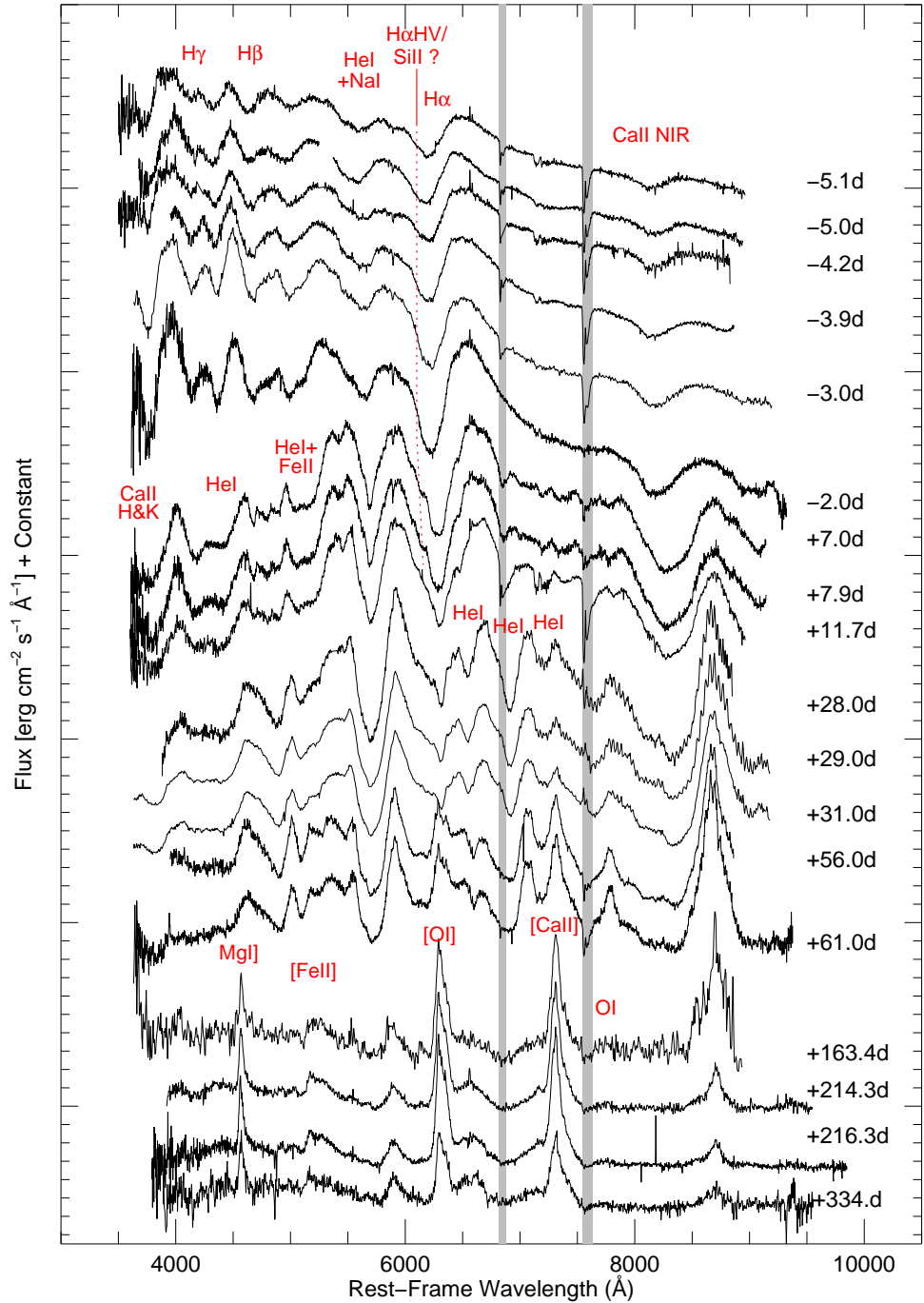}
\caption{Spectroscopic sequence of SN\,2011hs ranging from $-5.1$ to +334.0 days after $t(B)_{max}$. Each spectrum has been reported to the host galaxy rest frame, corrected for Milky Way and host galaxy reddening and, finally, shifted  by an arbitrary amount for presentation. On the right, the phase from $t(B)_{max}$ are labeled. Grey bands indicate the telluric bands position.  }
\label{spec}
\end{center}
\end{figure*}


\section{Spectral Evolution} \label{SpectralEvolution} 
\begin{figure*}
\begin{center}
\includegraphics [width=11cm,angle=90]{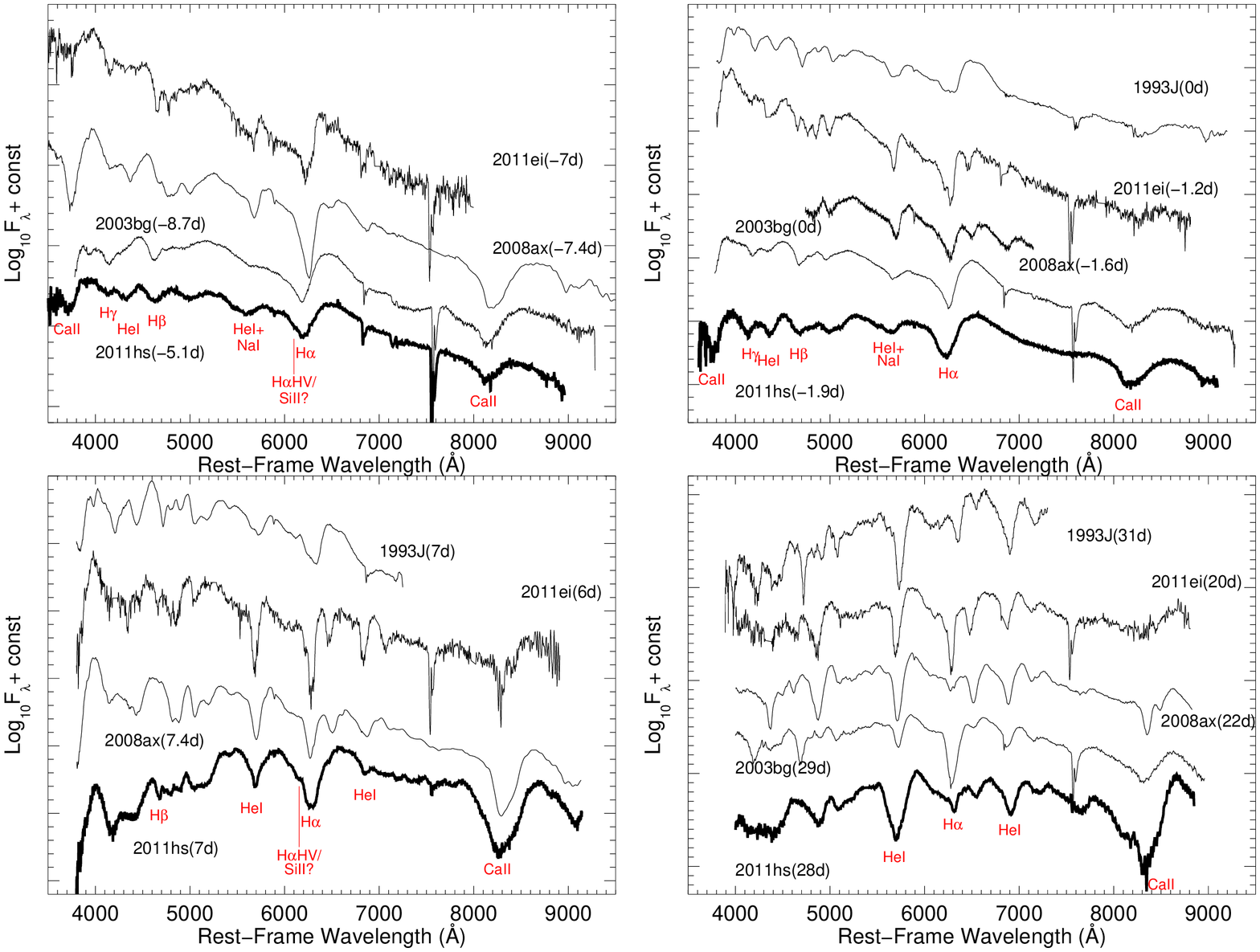}
\vspace{0.4cm}
\caption{Comparison of SN 2011hs spectra with those of Type IIb SNe 1993J (\citealt{Barbon93J}; \citealt{Richmond93J}), 2008ax (\citealt{Pasto08ax}; \citealt{Taubix08ax}), 2011ei \citep{Dan11ei} and  2003bg \citep{Hamuy03bg}, at  similar phases after $t(B)_{max}$, namely around $-$7 days ({\it upper left panel}), 0 days ({\it upper right panel}), 7 days ({\it lower left panel}) and 30 days ({\it lower right panel}). Spectra have been  corrected for total (Milky Way + host galaxy) reddening and  shifted to the host galaxy rest frame. }
\label{comp_spec}
\end{center}
\end{figure*}

Fig. \ref{spec} contains the spectral sequence of SN\,2011hs, ranging from about 5 days before $t(B)_{max}$ to about one year later (334\,days). At the earliest phases, the spectra are dominated by the H Balmer series lines, with the typical P-Cygni profile.
From the minimum of the absorption components, we measured an expansion velocity around 17,000\,km\,s$^{-1}$ for H$\alpha$, and 14,000\,km\,s$^{-1}$ for H$\beta$ and H$\gamma$, with the latter being slower because of a higher optical transparency of the ejecta at these wavelengths. 
Absorption features at $\sim$4300 \AA\ and $\sim$5600 \AA\ are identified as He\,I lines $\lambda\lambda$4472 and 5876, respectively,
with the latest likely blended with Na\,I.  
The broad absorption feature at $\sim$8200\,\AA\, is  due to the Ca\,II NIR triplet. Ca\,II H\&K absorption at around 3750 \AA\ is also present.  
 Evolving through maximum, the strength of H$\alpha$ changes, as well as its profile, with the emergence of a shoulder in the blue-wing, clearly detectable around one week after maximum.
 As recently discussed by \citet{Hachinger12}, this line (see Figures \ref{spec} and \ref{comp_spec}) could be identified as Si\,II, moving at early phases at about 12,000\,km\,s$^{-1}$ in SN\,2011hs case, or as  H$\alpha$ emission from a high velocity  H bubble with a velocity of $\sim$20,000\,km\,s$^{-1}$. 
 In the latter case, we would expect to detect such  high velocity components also in the H$\beta$ and H$\gamma$.
 Indeed, while near H$\gamma$ the S/N is too low for a detailed analysis, H$\beta$ seems to show a hint of  a double minimum 
 profile in the spectra between -4 and -3 days from maximum  with a possible high velocity component
 at $\sim$14,500\,km\,s$^{-1}$. However, we cannot discard the identification of these features with other ions, i.e. Co\,II or Fe\,II,
 as modeled by \citet{Mazzali03bg} and, recently, by \citet{Hachinger12}.\\
A week after maximum, the spectrum is dominated by strong absorptions of He I, with the features at around 6550, 6900 and 7200 \AA\ becoming more conspicuous and identified as He\,I $\lambda\lambda$6678, 7065 and 7281, respectively.  These lines  become dominant within one month after the maximum, whilst in the blue part of the spectrum, Fe\,II  lines emerge.\\
We followed the spectral evolution of the SN until its Sun conjunction, two months after $t(B)_{max}$.  When the SN became visible again, we obtained a nebular phase spectrum (+163 day), which displays  the typical emission features of a SE SN,
i.e. prominent Mg\,I]  $\lambda$4570, [O\,I] $\lambda$5577, [O\,I] $\lambda$$\lambda$6300, 6363 and  [Ca\,II] $\lambda$$\lambda$7291,7324 emission lines.
The [Fe\,II] emission at $\sim$5200\,\AA\ appears faint, suggesting that a small amount of $^{56}$Ni is produced in the explosion.
 Finally a boxy feature redwards of the [O\,I]  line at $\sim$6600 \AA\  is present and strengthens with time.  Nebular line profiles  are discussed in Sects. \ref{nebular_sec}  and \ref{modello}.


\subsection{Spectral Comparison} \label{spec_comp}
\begin{figure}
\begin{center}
\includegraphics [width=8.cm,angle=90]{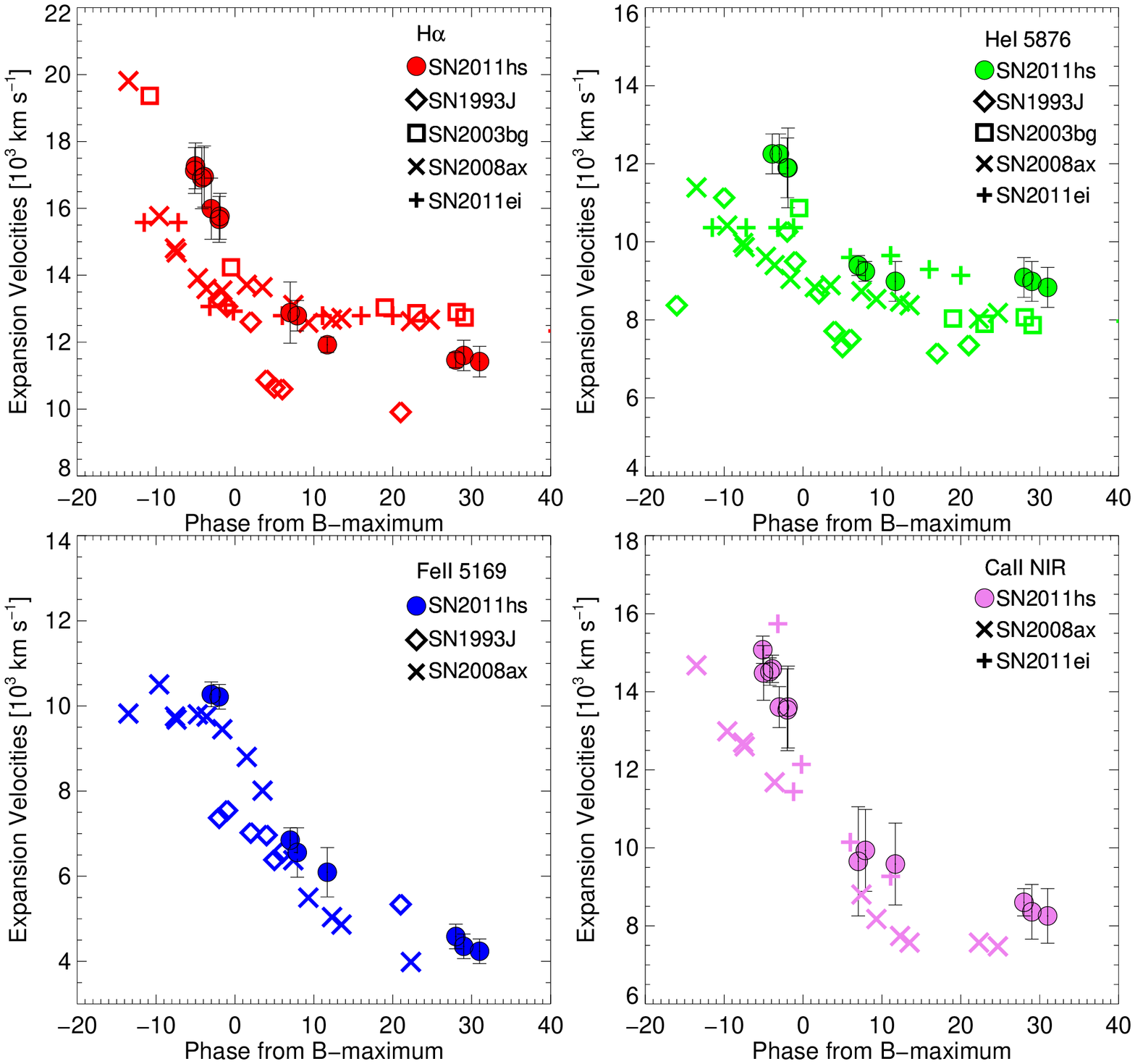}
\vspace{0.4cm}
\caption{Expansion velocities measured from the minimum of the absorption lines of the main ions in SN\,2011hs spectra.  
Velocities of the same ions in the spectra of the same SN sample of Figure\,\ref{comp_spec}. Uncertainties affecting each line velocity have been taken equal to three times the standard deviation from the  minimum position average. }
\label{velox}
\end{center}
\end{figure}

In Fig.\,\ref{comp_spec}, we compare SN 2011hs with other Type IIb SNe, namely SNe 1993J (\citealt{Barbon93J}; \citealt{Richmond93J}), 2008ax (\citealt{Pasto08ax}; \citealt{Taubix08ax}), 2011ei \citep{Dan11ei} and the fast expanding SN 2003bg \citep{Hamuy03bg}, at  similar phases after $t(B)_{max}$ (around $-7$ days, 0 days, 7 days and 30 days, respectively).
As discussed in the previous section, SN\,2011hs shows spectral features typical of Type IIb SNe but,
as clearly stands out from the comparison, they have broader profiles  and larger blue-shifts of the absorption minima, revealing higher  expansion velocities.
We determined the line velocities by fitting a Gaussian profile to their absorption features in the rest-frame spectra and measuring the blueshift of the minimum.\\ 
In Fig.\,\ref{velox}, we compare the time evolution of the expansion velocity of the most prominent lines (H$\alpha$, He\,I $\lambda$5876, Fe\,II $\lambda$5169 and Ca\,II NIR) of the same SN sample as Fig.\,\ref{comp_spec}.
In general, at similar epochs from $t(B)_{max}$, SN\,2011hs shows  higher velocities, 
similar only to those of SN 2011ei. 
Although having similar velocities, SNe 2011hs and 2011ei have very different line profiles, 
with the EWs of the H and He features in the spectra of SN 2011ei being among the narrowest ones for Type IIb SNe \citep{Dan11ei}.
The  spectral shape of SN 2011hs resembles more that of SN 2003bg: the two SNe show strong similarity
in most of the epochs in Fig.\,\ref{comp_spec}, with only one significant difference, namely 
the evolution of the H\,I lines that in SN 2011hs almost disappear a month after the maximum,
while in  SN 2003bg they remain conspicuous. This suggests a smaller H mass ejected by SN2011hs.
At the same epoch He I lines are more prominent  in SN 2011hs than in 2003bg,
pointing to a different distribution of the H and He in the ejecta or to a different degree of mixing (see \citealt{Taubix04aw}; \citealt{Hachinger12}).
In particular,  the SN\,2011hs H$\alpha$ expansion velocity declines more rapidly with time than SN\,2003bg, suggesting
a faster recession of the line forming region, possibly due to a lower density.
Most  importantly, SN 2011hs shows slightly higher Fe\,II velocity, where this ion is usually assumed as the best tracer
of the photospheric  expansion velocity. 
Thus the spectral comparison reveals a very fast expanding SN ejecta, i.e.  a high explosion 
energy per unit mass. 
The spectral similarity with SN\,2003bg (\citealt{Hamuy03bg}, \citealt{Mazzali03bg}) might suggest that SN\,2011hs is a Hypernova \citep{Hamuy03bg}, but its  fainter luminosity  and narrower light curve (as shown in Sec. \ref{lc_sec}) does not favor this interpretation. 
Finally, it is apparent from the different panels of Fig.\,\ref{comp_spec}, that there is an evolution of the  continuum shape of SN\,2011hs: initially similar to the other SNe (at -7 days), it becomes redder at 0 and 7 days past maximum, then returns to a similar colour after one month. Such behaviour is in agreement with the colour evolution found in Sect.\,\ref{color_sec}.


\subsection{NIR Spectroscoopy} \label{modello}
\begin{figure}
\begin{center}
\includegraphics [width=8.5cm]{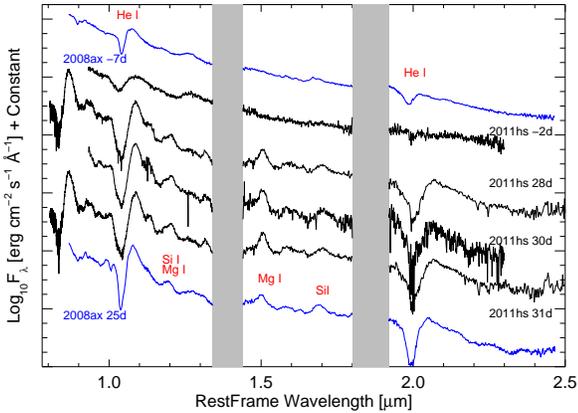}
\caption{NIR spectra of SN\,2011hs, compared with those of SN\,2008ax \citep{Taubix08ax} in pre-maximum and one month post-maximum phases. The spectra have been corrected for redshift and reddening with the same values as adopted for the optical data. Grey bands indicate the telluric bands position.}
\label{NIRevol}
\end{center}
\end{figure}

In Fig.\ref{NIRevol}, we show the four spectra we obtained  in the NIR wavelength range. They
basically cover two epochs of the SN spectral evolution, pre-maximum and 1-month after maximum phase.
The spectra obtained are precious, as only a few NIR spectroscopic data have been published to date of this class of SNe.\\
The $-2$ days spectrum is almost featureless, with exception of strong features at
$\sim1.03\mu$\,m and  $\sim1.22\mu$\,m.
The former may be associated to He\,I $\lambda1.083\mu$m with an expansion velocity of $\sim$13,800\,\,km\,s$^{-1}$.
Such a velocity is higher than those measured in the optical spectra taken at similar epochs, that may suggest a contamination
by other ions, e.g.  Mg\,II and/or C\,I. 
The presence of  C\,I can be probably excluded because then we would expect other strong lines from the same ion.
In general, He lines occur in spectral regions strongly affected by different metal absorptions  \citep{Lucy}, with exception of 
the He\,I $\lambda2.058\,\mu$m line, which has been claimed as the only direct evidence of the presence of this ion
 (\citealt{Hamuy02}; \citealt{Valenti07gr}; \citealt{Modjaz08D};  \citealt{Max07Y}).
For SN\,2011hs, there is a clear detection of the He\,I line at $\lambda2.058\,\mu$m: marginally detected at $-2$d, a strong P-Cygni emerges one month after maximum (see Fig.\ref{NIRevol}).\\
It is possible that Pa$\gamma$ absorption contaminates the $\sim1.03\mu$m feature in the $-2$ days spectrum, 
supported by the identification of the minimum at $\sim1.22\,\mu$m as Pa$\beta$, with an expansion
velocity of  $\sim$13,900\,\,km\,s$^{-1}$, similar to that measured for the optical H$\beta$ and H$\gamma$ at the same epoch
(Fig. \ref{velox}).  Unfortunately, Pa$\alpha$ lies in a region where the EarthÕs atmosphere is opaque,
so it cannot be detected.\\
In Fig.\ref{NIRevol},  we compare SN\,2011hs to  SN\,2008ax  \citep{Taubix08ax} at similar epochs.
For SN\,2008ax, a significant contribution of the H\,I Paschen lines   to the He\,I $\lambda1.083\,\mu$m absorption feature  has been excluded, because of the weakness of the Pa$\beta$ line already in the pre-maximum spectrum.
 At later phases,  we can recognize in SN\,2011hs the emergence of an emission band at $\sim1.19\,\mu$m likely attributed  to Si\,I $\lambda\lambda1.198,1.203\,\mu$m blended with Mg\,I $\lambda1.183\,\mu$m. The latter contributes also to the emissions at $\sim1.50\,\mu$m and at $\sim1.58\,\mu$m.
The O\,I $\lambda0.926\,\mu$m and  O\,I $\lambda1.315\,\mu$m are  responsible for the emission bands at $\sim0.93\mu$m and 1.31$\mu$m, respectively. 


\begin{figure}
\begin{center}
\includegraphics [width=0.48\textwidth]{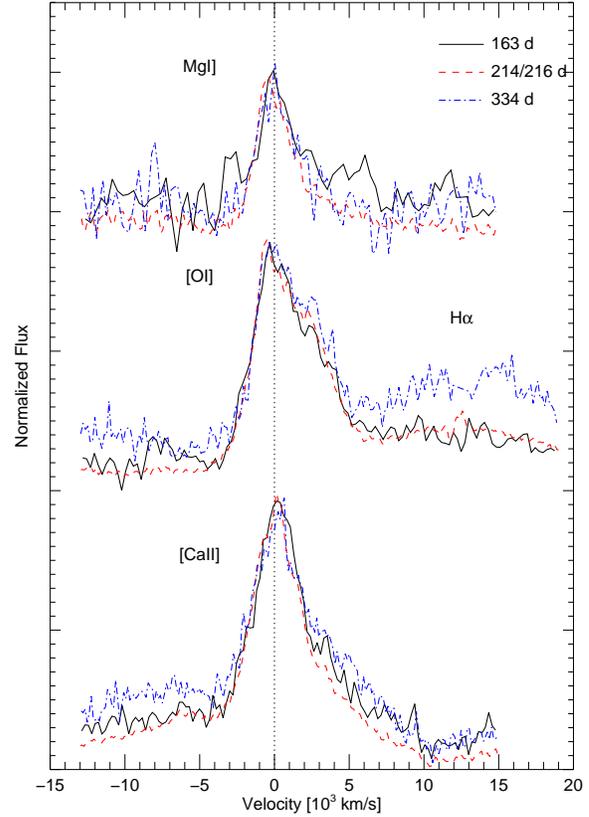}
\caption{Nebular emission line profiles of SN 2011hs from 163 days, 214/216 days and 334 days spectra.  }
\label{nebular}
\end{center}
\end{figure}


\subsection{Nebular Spectral Line Profile}\label{nebular_sec}
The profiles and velocities of nebular lines can provide valuable information concerning the  geometry of SN 
explosion and the distribution of the emitting material (Fransson \& Chevalier 1987, 1989; \citealt{Mazzali05}; \citealt{Maeda08}).  
Plotted in Fig. \ref{nebular} are the emission line profiles, in velocity space, of the predominant ions Mg\,I]  $\lambda$4570, [O\,I] $\lambda$$\lambda$6300, 6363 and  [Ca\,II] $\lambda$$\lambda$7291,7324, from spectra taken at 163, 214, 216 and 334 days after $t(B)_{max}$ (see Tab. \ref{spec_journal}).
Since no major evolution is detectable between them, the two spectra taken at +214d and +216d  have been combined to improve the signal-to-noise (indicated as 214/216 days). 
The nebular line profiles reveal similar widths and no evidence for asymmetries, suggesting a fairly 
symmetric expanding ejecta. 
A strong box-like emission profile red-ward of the [O\,I] doublet line is also present and identified as H$\alpha$ emission line.
Such  feature has been detected in previous SE SNe, e.g.  SN1993J \citep{Patat93J},  SN 2007Y  \citep{Max07Y} and SN 2008ax \citep{Taubix08ax},
and claimed to be the product of the interaction  between a fast expanding shell of H from the SN and the dense CSM  \citep{Chavalier94}.
This explanation raised some inconsistencies   between  the shock interaction  scenario
and the low H$\alpha$ velocity observed in e.g. SN 2008ax \citep{Taubix08ax}.\\
An alternative  interpretation, given by \citet{Maurer}, explains such late phase H$\alpha$  emission 
as the result of  mixed and strongly clumped H and He.  In this case, radioactive energy deposition can power H$\alpha$ completely without any need for an additional source of energy.  Clumpiness can significantly increase the relative strength of H$\alpha$, and combined with the mass of He or H in such
mixed fraction, makes possible to reproduce the spectra in several combinations (see \citealt{Maurer}). 
Although this prevents us from determining the H/He mass involved,
it can explain the different observed shapes in late H$\alpha$  emission profiles: strong and box-shaped, like in SNe1993J , 2007Y and 2008ax; or  weak emissions,   like for SNe 2001ig and 2003bg \citep{Maurer}.
Most importantly,  it solves the contradictions with the X-ray observations, when the shock interaction is assumed to be  responsible for the H$\alpha$ emission (see e.g. \citealt{Chevalier10}). 
In SN 2011hs the H$\alpha$ emission has a velocity of $\sim$6,000$-$8,000\,km\,s$^{-1}$ measured at the edge of the observed feature, similar to that found by \citet{Taubix08ax} for SN 2008ax and lower than that  found for SN 2007Y  (around 9,000$-$11,000 \,km\,s$^{-1}$;  \citealt{Max07Y}).\\
Finally, a bump blue-ward of the [Ca\,II] line is also visible, as previously seen in the nebular spectra of SNe 2008ax \citep{Taubix08ax} and 2011ei \citep{Dan11ei}  and probably due to a blended emission from He\,I $\lambda$7065 and  [Fe\,II]\,$\lambda7155$. \\
Fransson \& Chevalier (1987, 1989) have shown that the [Ca\,II]/[O\,I] ratio is a sensitive tracer of the core mass,
giving useful insight into the main sequence progenitor mass. Staying relatively constant at late phases, the ratio decreases for more massive cores.  
In SN 2011hs, the  strength of [Ca\,II] $\lambda$$\lambda$7291,7324 is comparable  to that of  [O\,I] $\lambda$$\lambda$6300, 6364, with a mean ratio $\sim$1.2 measured  from the three nebular spectra.
As a reference, such ratio was  about 0.5 in SN\,1998bw,  which was found to have a progenitor with a mass  $\approx$40 M$_\odot$ \citep{Iwamoto98bw}. 
For SNe 2007Y and 1993J   a ratio $>$0.5 was measured,  suggesting a progenitor star with a main sequence mass much lower than the progenitor of SN 1998bw, i.e. $\loa$ 20 M$_\odot$ (\citealt{Max07Y}; \citealt{Nomoto93J}; \citealt{Podsiadlowski93}).
For SN 2008ax, \citet{Taubix08ax}, measuring a  [Ca II]/[O I] $\sim$0.9,  claimed 
 a   low-mass progenitor in a binary system, rather than a single massive WR star.
Thus having a  line flux ratio slightly higher than that of SN 2008ax,  a low-mass progenitor star can be proposed for SN\,2011hs, too.

\subsection{Nebular Spectrum Modelling}\label{nebular_mod}
\begin{figure}
\begin{center}
\includegraphics [width=0.48\textwidth]{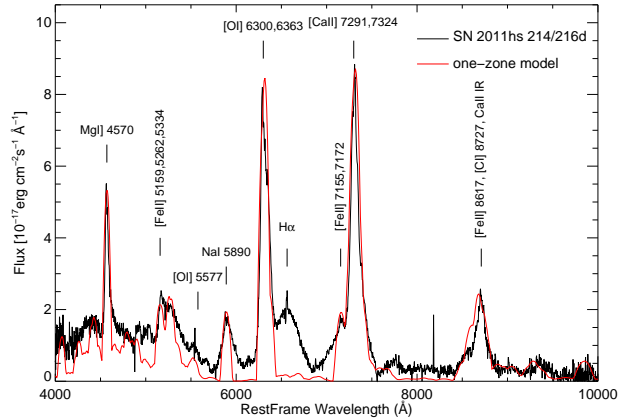}
\caption{The nebular spectrum of SN 2011hs taken at 214/216 d compared with the synthetic spectrum.  \label{syn_spec}}
\label{neb_fit}
\end{center}
\end{figure}

In order to establish the 
properties of the inner ejecta  with higher accuracy, we have modelled the 214/216d nebular spectrum.
Unfortunately due to the lack of  simultaneous photometric observations (or at least close in time),  the flux calibration of
 the 334d nebular spectrum is uncertain, thus it cannot be used for modelling.
We used our non-local thermodynamic equilibrium (NLTE)  code (e.g.
\citealt{Mazzali02ap}). The code computes the emission of gamma-rays and
positrons from the decay of $^{56}$Ni into $^{56}$Co and hence $^{56}$Fe. These are then
allowed to propagate in the SN ejecta, and their deposition is followed with a
MonteCarlo scheme as outlined in \citet{Cappellaro97}. Energy deposited heats
the gas, which is a mixture of various elements seen in the SN ejecta, through
collisional processes. Heating is balanced by cooling via line emission. The
balance of these two processes is computed consistently with the occupation of
the atomic energy levels, in NLTE. Emission is mostly in forbidden lines,
although some permitted transitions are also effective in cooling the ejecta.
Although the code is available in both a one-zone and a stratified version, here
we use the one-zone approach, because we do not have a viable explosion model
available for SN\,2011hs. 
Observed and synthetic spectrum are plotted in Fig.\,\ref{syn_spec}. We can establish from the line profiles a typical line
width of 3500 km\,s$^{-1}$. This matches the width of most emission features, except for
H$\alpha$, which is caused by a different mechanism  and is broader, reflecting the distribution of H.
Interestingly, H$\alpha$ does not develop a boxy profile, possibly indicating some
mixing of H down to low velocities. We find that a reasonable fit to the
spectrum (shown in Fig.\,\ref{syn_spec}), excluding H emission and the He zone, requires a small $^{56}$Ni mass,
0.04 M$_\odot$. The ejecta mass included within the boundary velocity is also quite
small, 0.23 M$_\odot$, as expected given the rapid evolution of the SN light curve.
These values are reasonably consistent with a He core of 3-4 M$_\odot$ found with the modeling of the bolometric light curve
(cf. Sect.\,\ref{bolom_mod}). The most abundant
element in the inner ejecta is Oxygen, as usual, with a quite small mass of 0.13 M$_\odot$, consistent with 
a low-mass star core-collapse scenario \citep{Limongi} and what we found in Sections\,\ref{nebular_sec} and \ref{bolom_mod}. Other elements that are seen in emission are C, Mg, Ca, Fe and Na.

\section{SN 2011hs physical parameters}\label{bolom_sec}

When no pre-explosion images are available, one of the most direct ways to estimate the physical parameters of a SN progenitor is by 
comparing the observations with the models of the light curve and the photospheric velocity evolution of the SN.  
The observed quantities adopted in the present work are a) the bolometric light curve (see Sect.\,\ref{bolom_lc}),
derived from the broad band photometry  and b) the photospheric velocity measured from Fe\,II lines. 
For the models, we used a one-dimensional, Lagrangian code described in \citet{Bersten_mod}.


\begin{figure*}
\begin{center}
\includegraphics [width=12cm,angle=90]{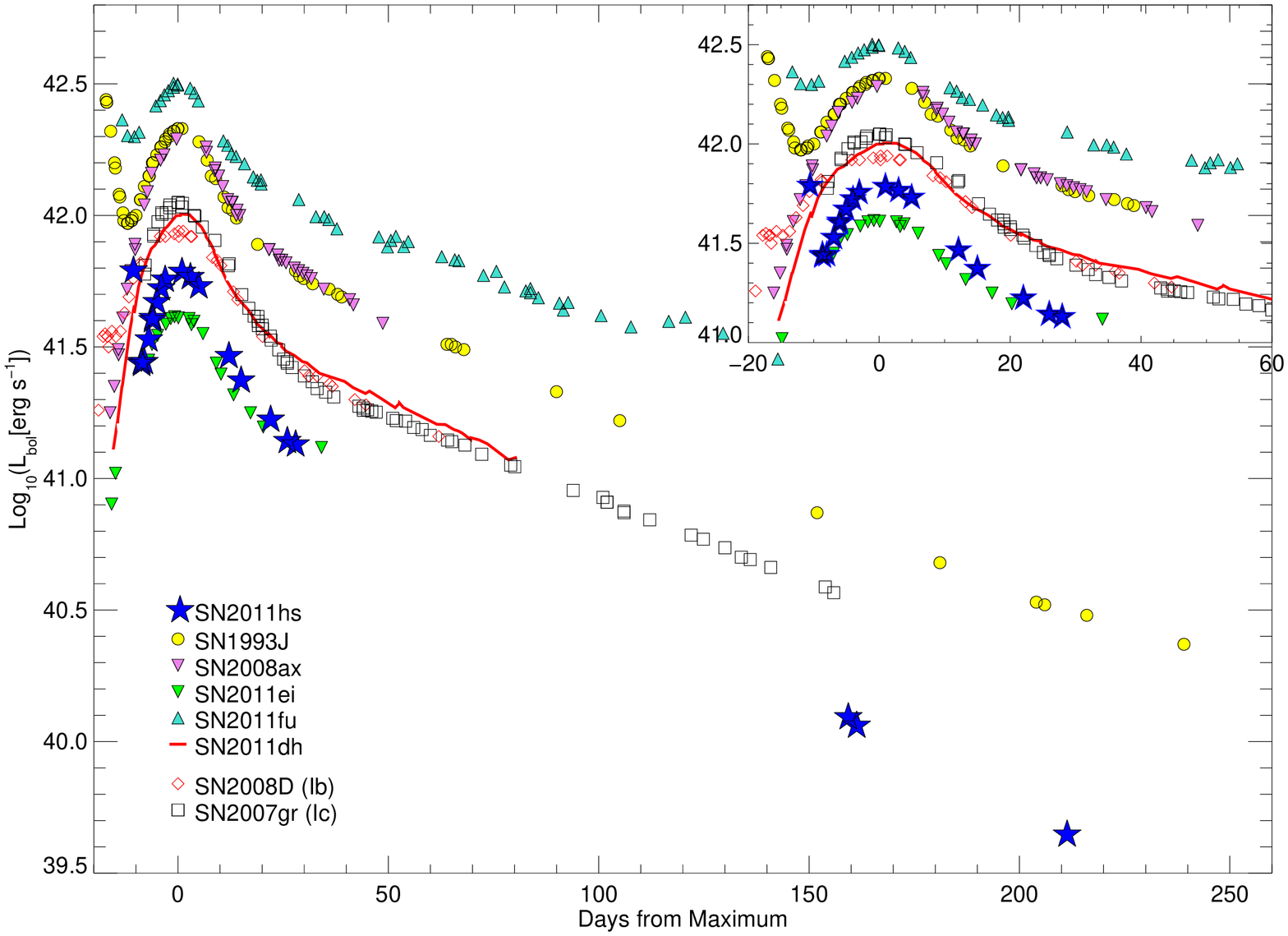}
\caption{ {\it BVRI} pseudo-bolometric light curve  of SN\,2011hs compared to those of the Type IIb SNe\, 1993J (\citealt{Barbon93J}; \citealt{Richmond93J}), 2008ax (\citealt{Pasto08ax}, \citealt{Taubix08ax}), 2011dh \citep{Ergon11dh}; 2011ei, \citep{Dan11ei} and 2011fu, \citep{Kumar11fu},
and the Type Ib SN\,2008D (\citealt{Mazzali08D}, \citealt{Tanaka08D}) and Type Ic SN\,2007gr \citep{Valenti07gr}. A zoom-in on the early photospheric evolution is given in the upper-right corner.}
\label{bolom}
\end{center}
\end{figure*}


\subsection{Bolometric  Light Curve}\label{bolom_lc}
To construct the bolometric light curve, we first corrected  the magnitudes for extinction and converted them to flux densities at the effective wavelength of the corresponding bandpass. The  total flux was then obtained by integrating  over the UV-opt-NIR wavelength range and, next, the integrated bolometric flux was converted into luminosity using the adopted distance (Section \ref{galaxy}). 
The  bolometric luminosity was computed for all the epochs by keeping the $R$ band light curve as reference and extrapolating
the missing data in other bands by assuming a constant colour.\\
We estimated the UV contribution to the bolometric emission  to be  around  20 per cent at very early phases and 
below 5\%  after maximum light. On the other hand, the NIR contribution did not exceed  30\% before maximum, while increasing to 50\% after.
Since a comparison of SN 2011hs with previously studied SE SNe would require making 
strong assumptions on the UV and NIR contributions to the bolometric flux for those missing  
data at these wavelengths, in order to be conservative, we compare their pseudo-bolometric flux, obtained by integrating only the {\it BVRI} light curves.
Thus, we compiled and compared the pseudo-bolometric curve of SN\,2011hs with those of the Type IIb SNe\, 1993J (\citealt{Barbon93J}; \citealt{Richmond93J}), 2008ax ( \citealt{Pasto08ax}, \citealt{Taubix08ax}), 2011dh \citep{Ergon11dh}; 2011ei \citep{Dan11ei} and 2011fu, \citep{Kumar11fu}, the Type Ib SN\,2008D (\citealt{Mazzali08D}, \citealt{Tanaka08D}) and Type Ic SN\,2007gr \citep{Valenti07gr}.
The comparison is displayed  in Fig.\,\ref{bolom}.
After SN\,2011ei, SN\,2011hs has one of the faintest bolometric luminosities at peak, with  L$_{BVRI}$= 6.1$\times$10$^{41}$ erg\,s$^{-1}$.
This indicates a smaller amount of $^{56}$Ni ejected with respect to the previous SE SNe: indicatively 
we can expect a total $^{56}$Ni mass between 0.03\,M$_\odot<$ M$_{Ni}<$ 0.09\,M$_\odot$, corresponding to the 
masses ejected by SN\,2011ei \citep{Dan11ei} and SN\,2008D \citep{Mazzali08D}, respectively.
The $^{56}$Ni mass is not the only factor that determines the light curve shape; the explosion energy and the ejected mass play an important role, too.
SN\,2011hs has a narrower light curve than the other SE SNe. 
Considering  Arnett's relation (\citealt{Arnett82}, 1996) for which $\tau_{\rm peak} \propto M_{\rm ej}^{3/4}E^{-1/4}$ and assuming a similar explosion energy, such narrow light curve width points to a smaller ejected mass, and, therefore, a smaller progenitor mass.

\subsection{The Light Curve modelling}\label{bolom_mod}
To  calculate models for the bolometric light curve  and the photospheric velocity 
evolution of SN 2011hs,  we used a code that  solves the hydrodynamics and
radiation transport  in an expanding ejecta including the gamma-ray transfer in grey approximation \citep{Bersten_mod}. 
He-core stars with different masses calculated from a single stellar evolutionary code \citep{Nomoto_mod} are used as the initial configurations
to explode \citep{Tanaka08D}. The initial density structures are artificially modified by attaching
a thin H-rich layer to the He core as required for a Type IIb classification. The method used to attach the envelope to the core was
recently presented in the modelling of  SN\,2011dh \citep{Bersten}. In this way it is possible to artificially modify the
progenitor radius and test it against observations.
As mentioned in Section\,\ref{lc_sec}, the light curve of a SE SN has two characteristic phases: (a) the early
UV/optical emission or ``cooling phase'' powered by the energy deposited by the shock wave and 
(b) a re-brightening to a broad maximum due to the decay of radioactive material  synthesized
during the explosion. Before rising to the light curve maximum, a minimum
or ``valley'' can be distinguished. While the global properties of a SN such as explosion energy ($E$), ejected mass
($M_{\mathrm{ej}}$), and the $^{56}$Ni mass can be obtained by
modelling the light curve around maximum,  observations of the early emission provide unique information
about the progenitor radius. In the case of SN\,2011hs, there is a single data point 
during the cooling phase that we used to place constraints on the possible progenitor radius.\\
 Firstly we compared the observables and the  theoretical bolometric 
 light curves and velocities focusing on the evolution  around the maximum light to obtain the general properties. 
  An important point required to reliably determine  physical parameters 
is the knowledge of the explosion time (t$_{\mathrm{exp}}$). Unfortunately, for SN\,2011hs  
there is no pre-explosion information available to help us  determine t$_{\mathrm{exp}}$, i.e. non-detection images
taken shortly before the explosion (see \citealt{CBET}),  hence, our estimate must rely on
comparisons with other Type IIb SNe.
 One possibility is to match the valley point of the light curve  of SN\,2011hs  with that of SN\,1993J,
 for which the explosion date and thus, the cooling duration is well known. In
 this case, in order to have an 8-day long cooling branch, t$_{\mathrm{exp}}$ should occur  $\sim$6 days before the  SN discovery,
 $\Delta$t = t$_{\mathrm{discovery}}-$ t$_{\mathrm{exp}}$= 6 days. 
 On the other hand, comparing the cooling in the $R$ band, we find a declining rate for SN\,2011hs  ($\Delta R$=0.46\,mag\,d$^{-1}$),
 which is more than twice that of SN 1993J ($\Delta R$=0.21\,mag\,d$^{-1}$, \citealt{Richmond93J}),
  implying a 4-day long branch (similar to that of SN 2011dh) and a $\Delta$t = 2 days.
  The latter case implies a rise  time to maximum light ($t(V)_{max}$) for SN~2011hs of $\approx$13 days, which is far from the
  observed typical value of $t(V)_{max}\approx$20 days  (\citealt{Richardson}; \citealt{Drout}).
  
 Alternatively,  in order to have a $t(V)_{max}\sim$20\,days after the explosion SN\,2011hs should have exploded
 with a   $\Delta$t = 9 days. Such a  t$_{\mathrm{exp}}$  would imply an extremely large radius  for a Type IIb SN progenitor star \citep{Bersten}, therefore
  we discard this scenario.\\
 Thus, in the following analysis, we adopt a $\Delta t=6$ days, assuming that  SN~2011hs exploded at $t_{\mathrm{exp}}=2,455,872 \pm 4$ JD, with a
  $t_{\mathrm{valley}} \sim 8$ days and  $t_{\mathrm{max},V} \sim 16$ days. 
This is in agreement with the explosion epoch obtained through the radio light curve modelling (see Section\,\ref{radio}).
Nevertheless, later in this section we further discuss the possibility of an earlier t$_{\mathrm{exp}}$.


\begin{figure}
\includegraphics[width= 6cm, angle=-90]{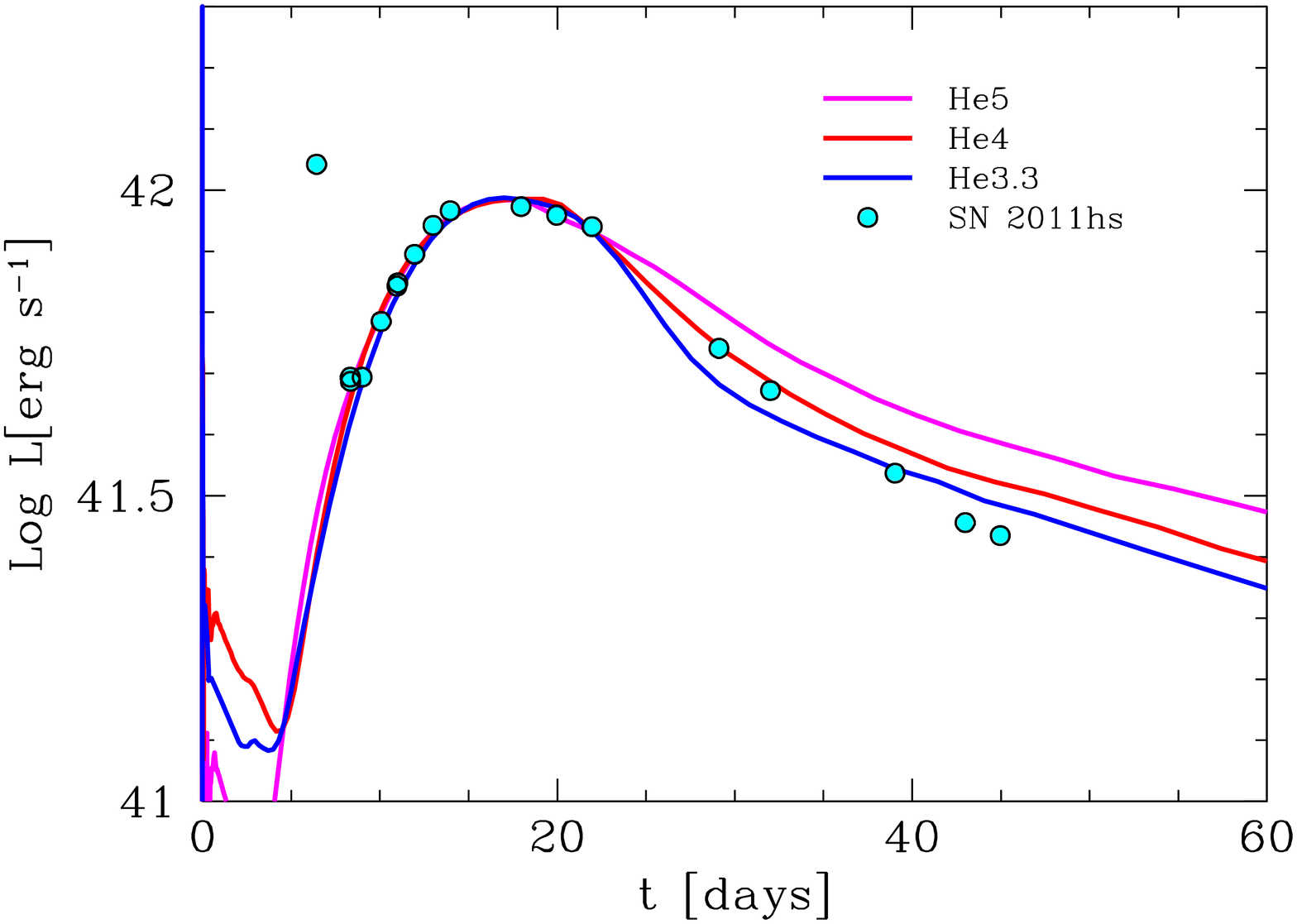}
\includegraphics[width= 6cm, angle=-90]{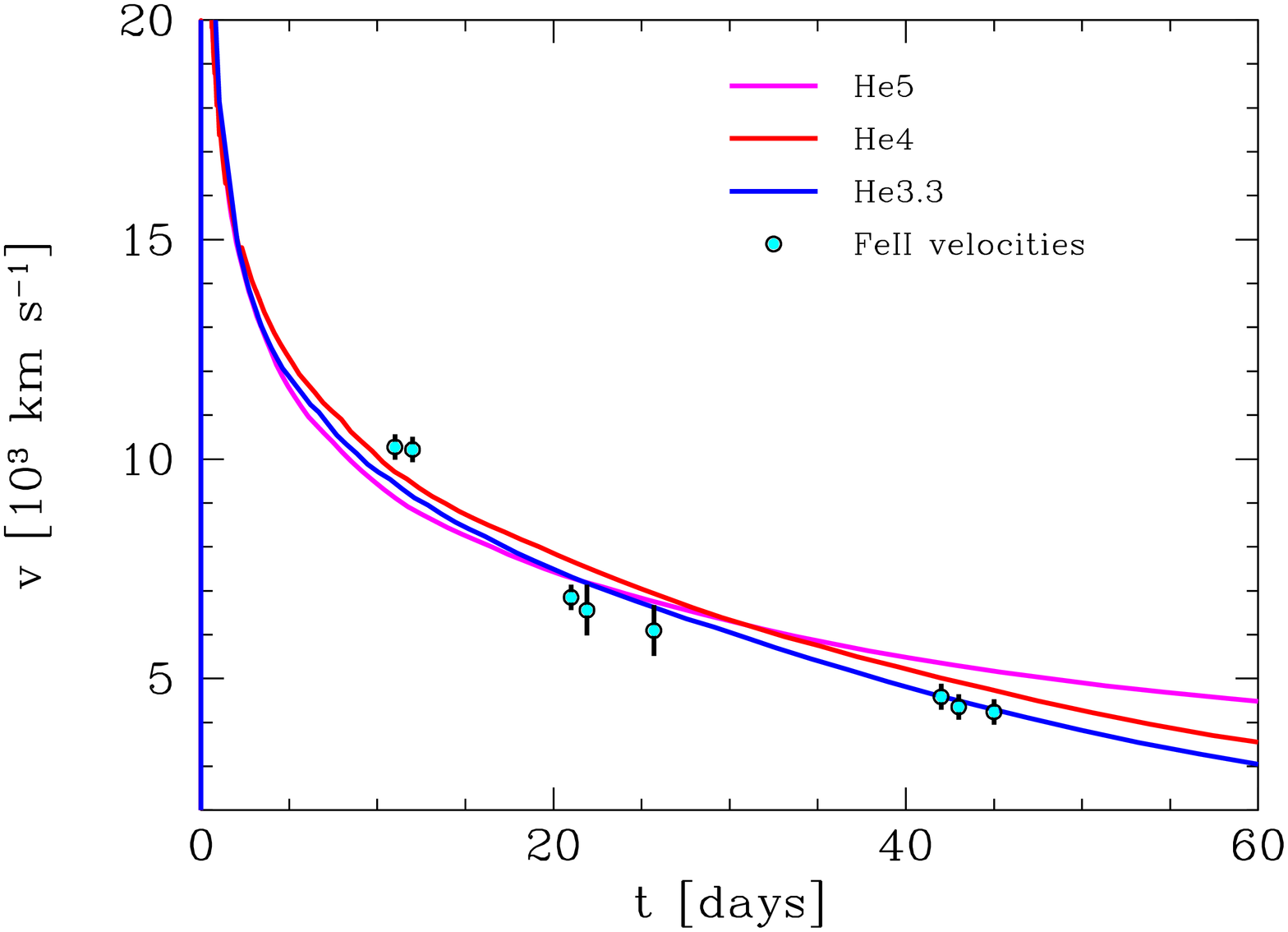}
\caption{{\em(Upper panel)} Observed bolometric light curve of SN
2011hs (points) compared with the results of the light curve
calculations for models He$3.3$ (blue line), He4 (red line) and He5 
(purple line). {\em(Lower panel)} Evolution of the photospheric
velocity for models He$3.3$ ( blue line), He4 (red line) and He5 (purple 
line) compared with measured Fe II line velocities of SN~2011hs. \label{Lbol:He4He3}}  
\end{figure}


We have calculated a set of models with different values of 
$E$, He-core mass and mixing of $^{56}$Ni to try to reproduce the
bolometric light curve around the main peak along with the photospheric 
velocity evolution. Specifically, we use three different pre-SN models
with He cores of $3.3$ $M_\odot$ (He$3.3$), 4 $M_\odot$ (He4) 
and 5 $M_\odot$ (He5), which correspond to the stellar evolution of
single stars with main-sequence masses of 12 $M_\odot$, 15 $M_\odot$
and 18 $M_\odot$, respectively. Fig.\,\ref {Lbol:He4He3} shows the best
results of such models (He$3.3$ in blue, He4 in red and He5 in purple
solid lines) for the bolometric light curve (upper panel) and
for the photospheric velocities (lower panel) compared with
the observations. Assuming  $M_{\mathrm{ej}}$= $M_{\mathrm{total}}-M_{\mathrm{cut}}$ (where
   $M_{\mathrm{cut}}$ is the mass of the compact remnant assumed to be
   $1.5$ $M_\odot$), the parameters used in each calculation are (a)
for He$3.3$: $E= 8 \times 10^{50}$ erg, $M_{\mathrm{ej}}=  1.8$ $M_\odot$ and a $^{56}$Ni mass of $0.037$
$M_\odot$; (b) for He4: $E= 9 \times 10^{50}$ erg,
$M_{\mathrm{ej}}= 2.5$ $M_\odot$ and $^{56}$Ni mass of $0.038$
$M_\odot$; and (c) for He5: $E= 1 \times 10^{51}$ erg,
$M_{\mathrm{ej}}= 3.5$ $M_\odot$ and $^{56}$Ni mass of $0.040$
$M_\odot$. In all cases the degree of $^{56}$Ni mixing assumed was
$\approx$ 80\% of the initial mass. From Fig.\,\ref{Lbol:He4He3}, we see that 
He$3.3$ and He4 models give a reasonably good match to the
observations, while He5 model, which can  also reproduce the light curve for $t
<$ 20 days,  clearly fails at later epochs. 
Since changing the physical parameters for such initial mass does not
improve the agreement with the observed data, we discarded models with He core mass $\ge$5\,$M_\odot$. 
The models He$3.3$ and He4 can be considered at the same level of
agreement with the data.


\begin{figure}
\includegraphics[width= 6cm,angle=-90]{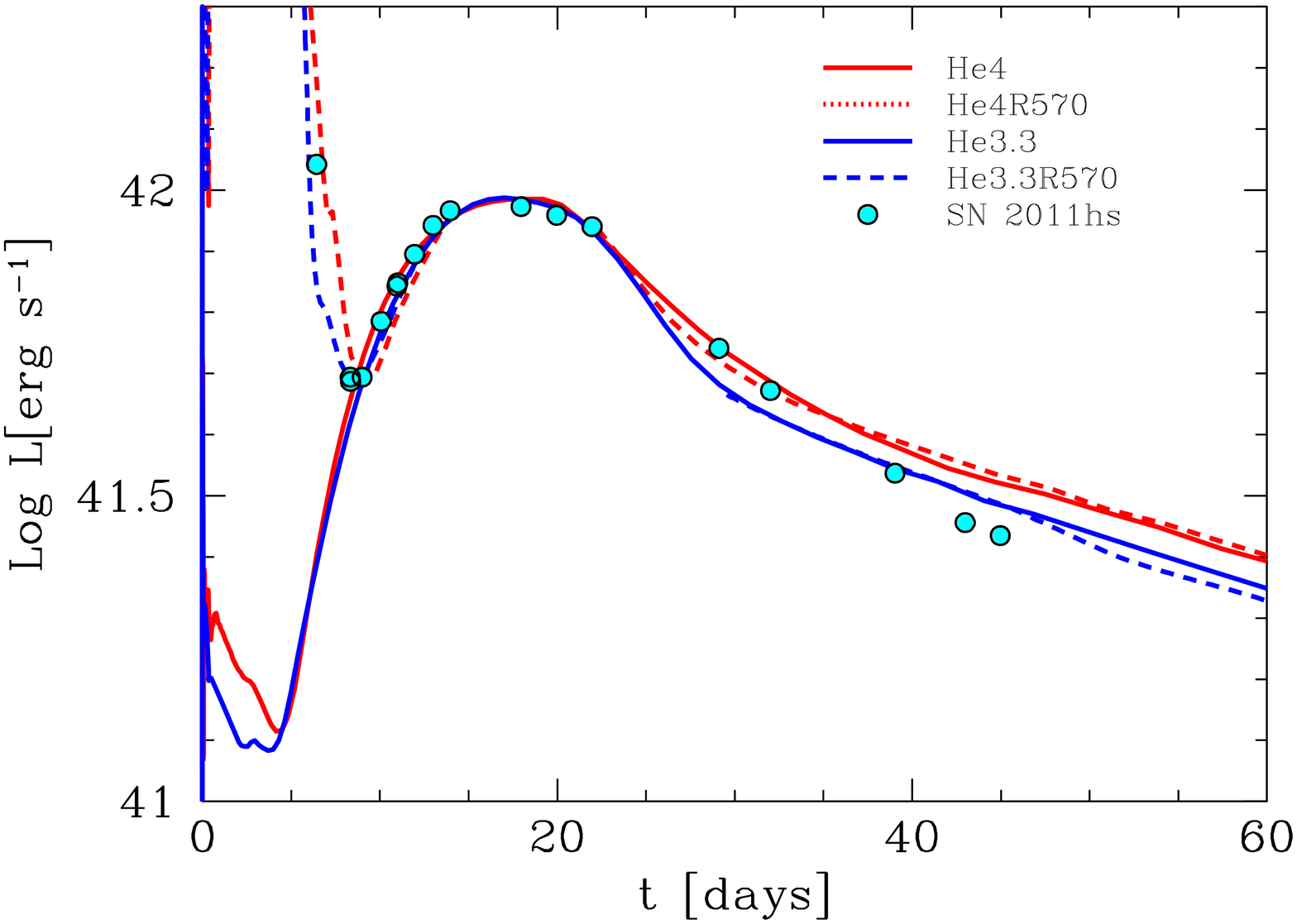}
\includegraphics[width= 6cm,angle=-90]{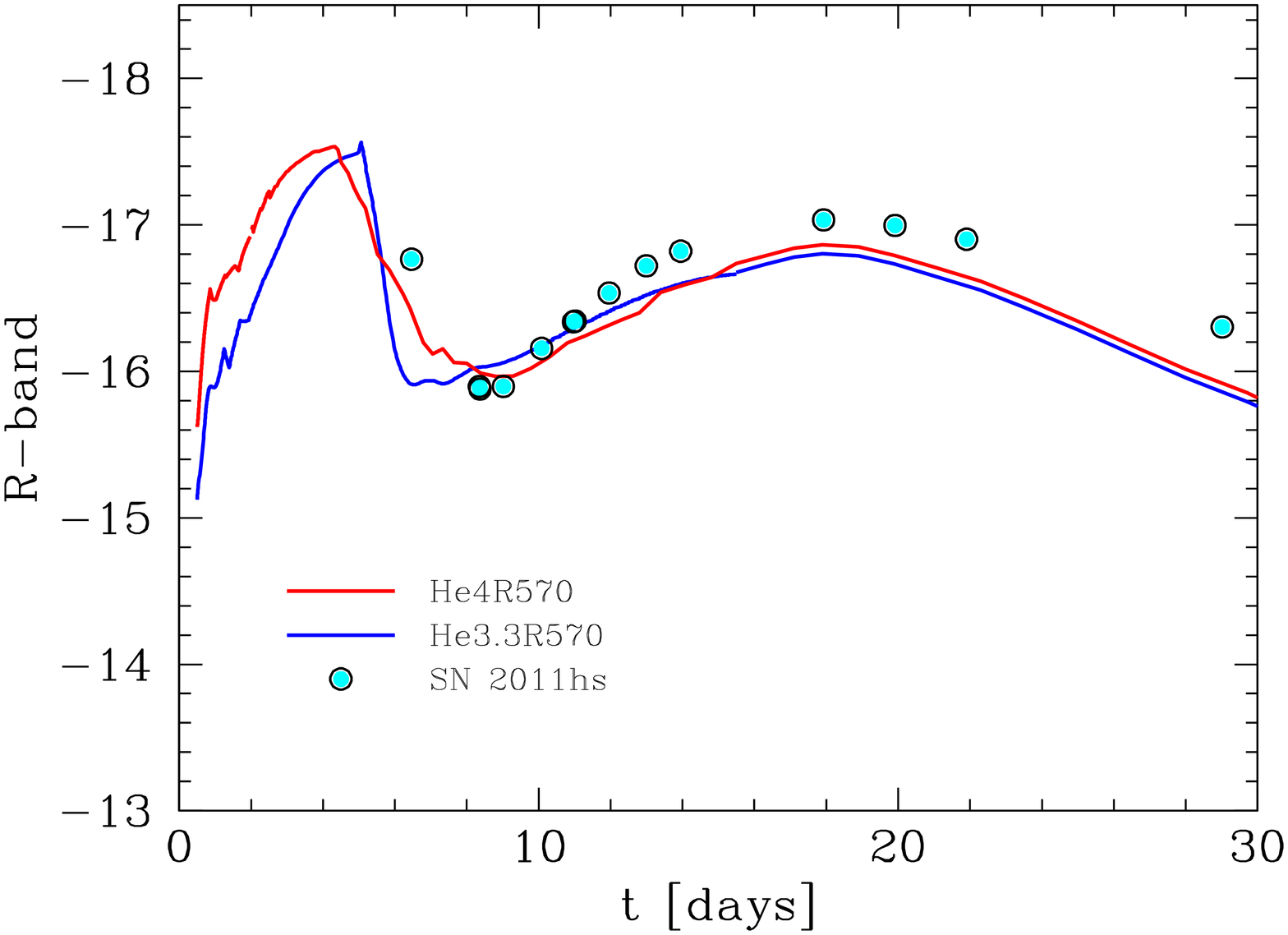}
\caption{\label{Lbol:radius}{ \em(Upper panel)}  Observed bolometric light curve
  of SN~2011hs (points) compared with the results of the light curve
  calculations for models He$3.3$ (blue line) and He4 (red
  line).  Two models with larger progenitor radii, He$3.3$R570
  (blue dot line), He4R570 (red dot line), are also shown (see text
  for details). Only with extended structures of $R > 500
  R_\odot$ it is possible to reproduce the earliest data point and the
  light curve valley. {\em (Lower panel)} $R$-band light curve for extended
  models He3R570 (blue line), He4R70 (red line) compared with the
  observations.}
\end{figure}


\begin{figure}
\includegraphics[width= 6cm, angle=-90]{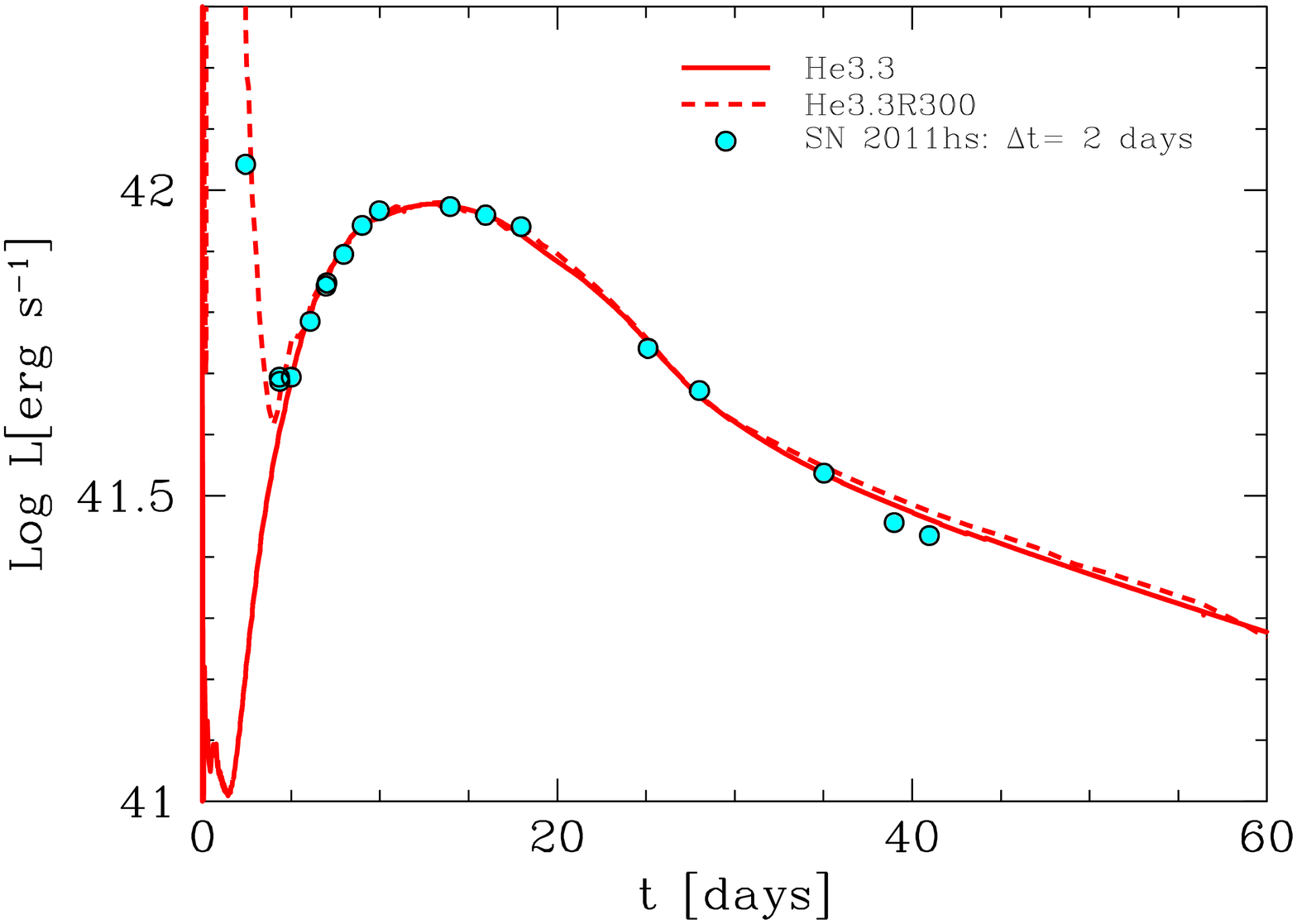}
\includegraphics[width= 6cm, angle=-90]{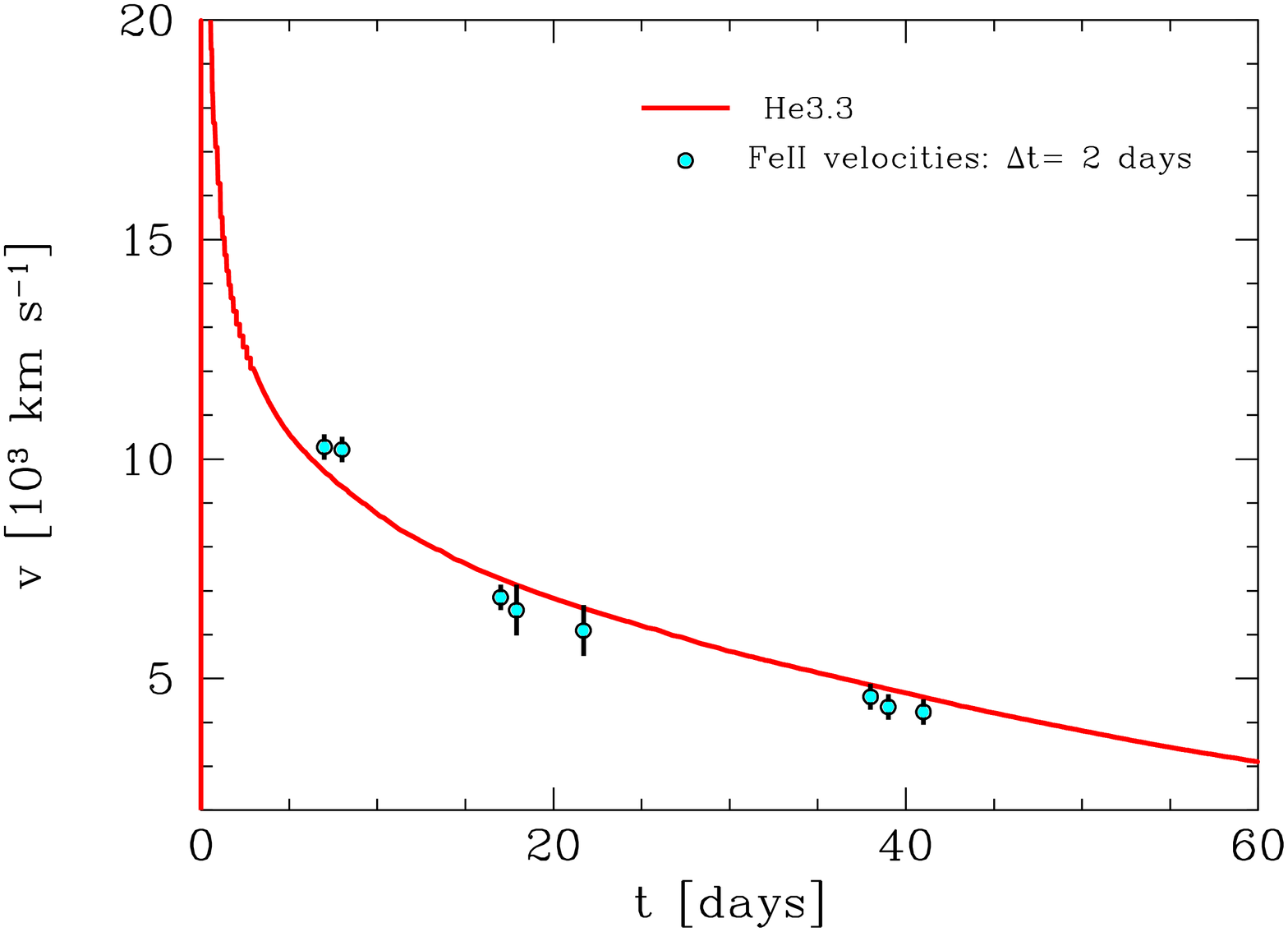}
\caption{\label{Lbol:He3} {\em (Upper panel)}  Observed bolometric light curve
  of SN~2011hs assuming the explosion occured two days before the SN
  discovery ($\Delta t=2$; cyan points) compared with the results of the light curve
  calculations for the He$3.3$ model with (dashed line) and
without (solid line) an attached H-rich envelope. {\em (Lower panel)}
Evolution of the photospheric velocity for models He$3.3$
compared with measured line velocities of Fe~II.} 
 \end{figure}

Both models, He3.3 and He4, have a compact structure with a radius $\approx$2\,R$_\odot$ and clearly cannot reproduce the earliest data point
as well as the luminosity of the valley observed in SN~2011hs. 
To improve  the fit during the cooling phase, we have attached several
envelopes to He3.3 and He4 with  different radii.
The best models we found were a model with a radius
of 570 R$_\odot$ attached to He3.3 (He3.3R570) and   to He4 (He4R570). The mass of the envelope we assumed for both models was $M_{\mathrm{env}} <
0.5M_\odot$. Models are shown in Fig.~\ref{Lbol:radius} (upper panel). 
Since we have no information on the colour evolution of SN 2011hs along the cooling decline,
we could introduce an uncertainty in the bolometric flux estimation with the adopted assumption of constant color  (see Sect.\,\ref{bolom_lc}).
Thus, we also compare SN\,2011hs $R$ band light curve with the same models found using the
bolometric one. Fig.\,\ref{Lbol:radius} (lower panel) confirms the good agreement of the models (especially for the He4 model) with the  
observations. However, we stress that in order to calculate the theoretical $R$ band light curve, we assumed 
a black body emission which may not necessarily be the case, especially at late epochs.\\
Given the  uncertainties affecting the observional data (extinction, t$_{\mathrm{exp}}$, etc) and the 
models (simple prescription of the radiation transfer, one
dimensional calculations, differences in the initial model from
different stellar evolutionary calculation, etc), the models
indicate a range of validity for the physical parameters of
SN~2011hs rather than robust estimations.
Our analysis suggests a progenitor star composed of a He core of 3--4 $M_\odot$ and a thin
H-rich envelope of $<0.5M_\odot$, for  a main sequence mass
estimated to be in the range of 12--15  $M_\odot$ (based on our stellar
initial model). 
To reproduce the early light curve of SN  2011hs, a progenitor radius in the range of 500--600 R$_\odot$
is required. 
An explosion energy of  $E\sim8.5 \times 10 ^{50}$
erg, a $^{56}$Ni mass of about 0.04 M$_\odot$ and a  mixing of 80 $\%$ of the initial mass  
reproduce well the observations around the light curve maximum.   
 Finally, note that our modelling  rules out progenitors with
He core mass $>5 M_\odot$, which excludes  main sequence masses above
$20$ $M_\odot$. For comparison, in the case of SN~2011dh initial
masses above 25 $M_\odot$ were ruled out (Bersten et al. 2012),  thus 
this possibly implies that the progenitor of SN 2011hs was less
massive than that of SN 2011dh.

\subsubsection*{Short rise time}
We analyze the possibility of
$\Delta t= 2$ days, although this value would imply a rise time to the peak for
SN~2011hs of $\approx$13 days, which is lower than the typical values
for SE SNe and in disagreement with the result of radio data modeling.
We have tested the same initial models as in the previous section,
i.e. He core masses of $3.3$ M$_\odot$ (He$3.3$), 4 M$_\odot$ (He4)
and 5 M$_\odot$ (He5). For models He4 and He5, we could not find a
set of parameters that can reproduce simultaneously the light curve and the
photospheric velocities. However, for the least massive model, He$3.3$, 
we found a very good agreement with the observations. Fig.\,\ref{Lbol:He3}  shows
this model with and without an attached envelope, compared
with the observations. The physical parameters used in this simulation
 are $E= 6 \times 10^{50}$ erg, $^{56}$Ni mass of 0.037 $M_\odot$,
an H-rich envelope with a radius of $\approx$ 300 $R_\odot$ and a
mass of $0.1$ M$_\odot$ (He$3.3$R300). However, we had to assume
almost complete mixing of $^{56}$Ni ($\approx$ 98\% of the initial
mass) to fit the rising part of the light curve. Note that the $^{56}$Ni mass
is the same as that 
found in the previous section, but in this 
case a less energetic explosion and a less
extended progenitor were needed. This analysis shows that the two parameters that
change most dramatically with the assumed explosion time are the
mixing and the progenitor radius. Therefore, it is important to know
$t_{\mathrm{exp}}$ as best as possible if one wants to predict these
parameters accurately.
We believe that there is no reason to assume such extreme
$^{56}$Ni mixing as there is no strong evidence of large asymmetries
in the explosion of SN~2011hs. Therefore we consider this fast
rise-time scenario as less likely than the one presented previously. 
However, note that even with this $t_{\mathrm{exp}}$ we had
to assume an extended object (300 $R_\odot$) to reproduce the earliest
data point.


\begin{figure*}
\rotatebox{270}{\includegraphics[width=12cm]{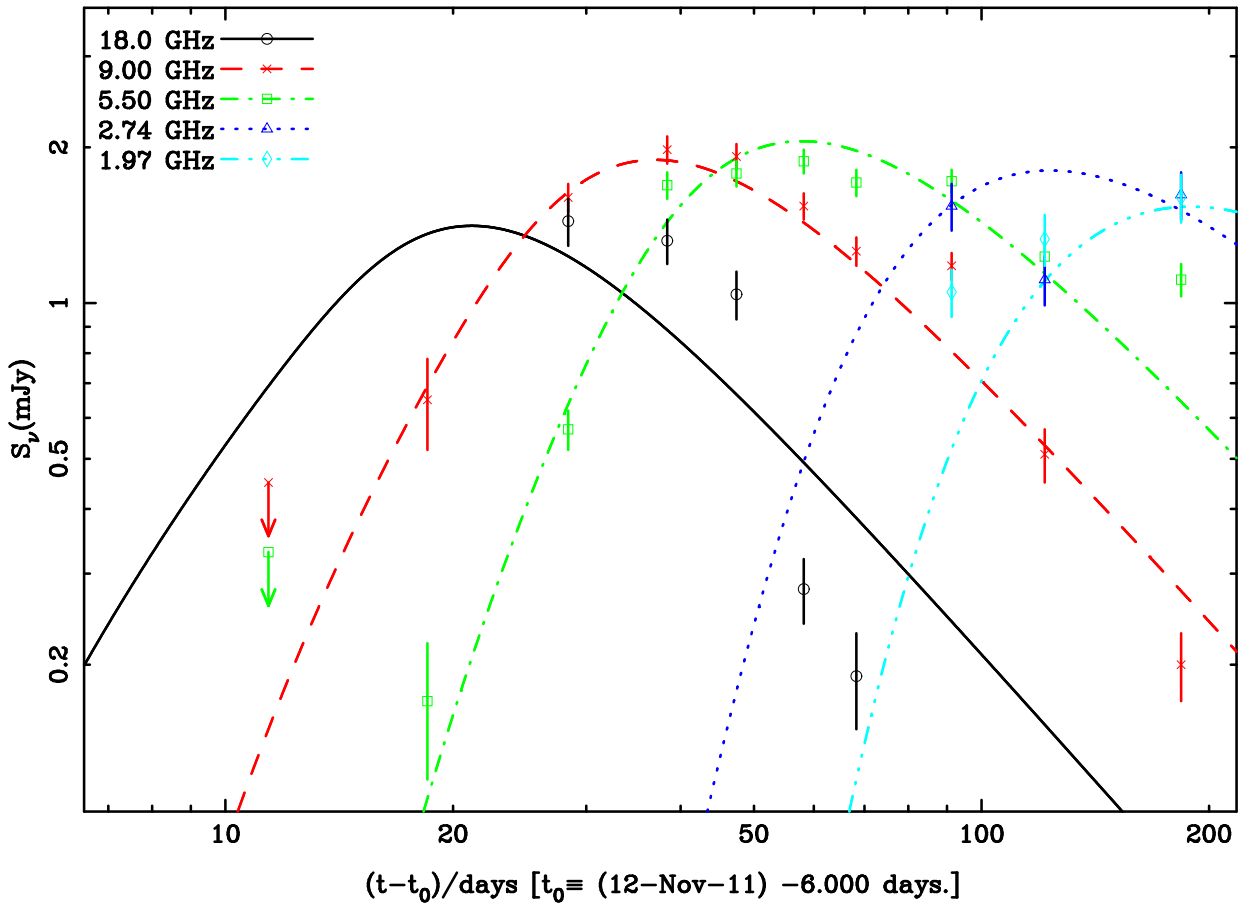}}
  \caption{\label{f:rlc} Radio light curves of SN 2011hs at  frequencies of 18.0~GHz
  ({\em black circles, solid line}), 9.0~GHz ({\em red crosses, dashed line}),
  5.5~GHz ({\em green squares, dash-dotted line}), 2.7~GHz ({\em blue triangles,
  dotted line}), and 2.0~GHz ({\em cyan diamonds, dash-triple dotted line}).
  The curves are a model fit to the data, as described in the text. }
\end{figure*}


\section{Radio data analysis}\label{radio}
Radio studies of SNe (RSNe) can provide valuable information
about the density structure of the circumstellar medium, the late
stages of stellar mass-loss, and clues to the nature of the progenitor
object \citep{kw02}. 
Radio emission has only been detected from
core-collapse SNe, and observed among these to date only from 12 Type~IIb SNe: 
SN~1993J \citep{kw07}, SN~1996cb \citep{Weiler96cb}, SN~2001gd
\citep{sto07}, SN~2001ig \citep{ryd04}, SN~2003bg \citep{sod06},
SN~2008ax \citep{sto08a,rom09}, SN~2008bo \citep{sto08b}, SN~2010P (\citealt{Herrero10P}; \citealt{Romero}), SN~2011dh
\citep{hor12,kra12,Soderberg12}, SN~2011hs \citep{ryd11}, PTF~12os
\citep{sto12} and the recent SN~2013ak  \citep{cha13}.\\
It has been found that the radio ``light curve'' of a core-collapse supernova can be broadly
divided into three phases. First, there is a rapid turn-on with a
steep spectral index ($\alpha>2$, so the SN is brightest at higher
frequencies) due to a decrease in the line-of-sight absorption. After
some weeks or months have elapsed, the flux reaches a peak, turning
over first at the highest frequencies. Eventually, the SN begins to
fade steadily, and at the same rate at all frequencies, in the
optically-thin phase.


\begin{figure}
\includegraphics[width=6.cm,angle=270]{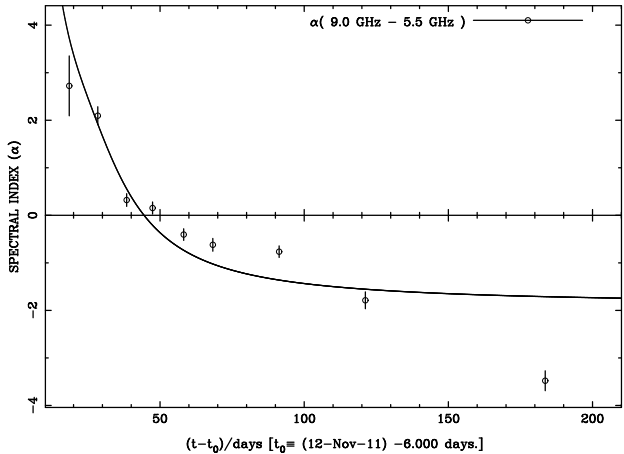}
  \caption{\label{f:spix}Evolution of spectral index $\alpha$ for SN~2011hs,
  plotted linearly as a function of time, between 5.5 and 9.0~GHz.
  The curve is a model fit to the data, as described in the text.}
\end{figure}

The ATCA radio light curve of SN\,2011hs is plotted in Figure~\ref{f:rlc},
 while the time evolution of the spectral index $\alpha$ (where flux $S \propto
\nu^{+\alpha}$) between simultaneous positive detections at 5.5 and
9.0~GHz is plotted in Fig.~\ref{f:spix}. \\ 
As can be seen in Fig.~\ref{f:rlc}, the radio light curves for
SN~2011hs are broadly consistent with the typical scenario described above. The emission had
already peaked at 18~GHz prior to the first observation at that
frequency; the peak at 9~GHz occurred about a month after discovery;
the peak at 5.5~GHz almost a month later; and the SN may just have
peaked at 2~GHz by the time observations ended. There are a few points
at each frequency which appear to exhibit significant departures from a
smooth evolution, but none are as achromatic or periodic as that
displayed by SN~2001ig or SN~2003bg. On the other hand SN~2011hs never
rose above a flux of 2~mJy at any frequency, which was about where
monitoring of SN~2001ig had to be suspended with the much less
sensitive ATCA$+ 2 \times 128$~MHz bandwidths.

The general properties of supernova radio light curves as outlined
above are quite well represented by a modified version of the
``minishell'' model of \citet{che82}, and have been successfully
parameterised for more than a dozen RSNe (see Table~2 of
\citealt{kw02}). Radio synchrotron emission is produced when the SN
shock wave ploughs into an unusually dense circumstellar medium
(CSM). Following the notation of \citet{kw02} and \citet{sw03}, we
model the multi-frequency evolution as:

\begin{eqnarray} 
S {\rm (mJy)} = K_1 {\left({\nu} \over {\rm 5~GHz}\right)^{\alpha}}
{\left({t - t_0} \over {\rm 1~day}\right)^{\beta}} e^{-{\tau_{\rm external}}} 
 \nonumber \\
\times
{\left({1 -e^{-{\tau_{\rm CSM_{clumps}}}} \over {\tau_{\rm CSM_{clumps}}}}
\right)} {\left({1 - e^{-{\tau_{\rm internal}}}} \over {\tau_{\rm internal}}
\right)}
\end{eqnarray}

\noindent
with 

\begin{equation}
\tau_{\rm external} = \tau_{\rm CSM_{homog}} + \tau_{\rm distant}
\end{equation}

\noindent
where

\begin{equation} 
\tau_{\rm CSM_{homog}} = K_2 {\left({\nu} \over {\rm 5~GHz}\right)^
{-2.1}}
{\left({t - t_0} \over {\rm 1~day}\right)^{\delta}},
\end{equation}

\begin{equation} 
\tau_{\rm distant} = K_4 {\left({\nu} \over {\rm 5~GHz}\right)^{-2.1}},
\end{equation}

\noindent
and

\begin{equation} 
\tau_{\rm CSM_{clumps}} = K_3 {\left({\nu} \over {\rm 5~GHz}\right)^{-2.1}}
{\left({t - t_0} \over {\rm 1~day}\right)^{\delta^{\prime}}},
\end{equation}

\noindent with the various $K$ terms representing the flux density
($K_1$), the attenuation by a homogeneous absorbing medium ($K_2$,
$K_4$), and by a clumpy/filamentary medium ($K_3$), at a frequency of
5~GHz one day after the explosion date $t_0$. The $\tau_{\rm
CSM_{homog}}$ and $\tau_{\rm CSM_{clumps}}$ absorption arises in the
circumstellar medium external to the blast wave, while $\tau_{\rm
distant}$ is a time-independent absorption produced by e.g., a
foreground H\,{\sc ii}~region or more distant parts of the CSM
unaffected by the shock wave. The spectral index is $\alpha$, $\beta$
gives the rate of decline in the optically-thin phase; and $\delta$
and $\delta^{\prime}$ describe the time dependence of the optical
depths in the local homogeneous, and clumpy/filamentary CSM,
respectively (see \citet{kw02} and \citet{sw03} for a detailed account
of how these parameters are related). For lack of sufficient
high-frequency data prior to the turnover to constrain it, we adopt
${\tau_{\rm internal}}=0$.

In order to assess the gross properties of SN~2011hs, we have fit this standard model to all the
data points plus upper limits in Table~\ref{t:radiofluxes}.  The
actual date of explosion $t_0$ is found to be {\it at least 5 days} prior to
discovery in agreement with the hydrodynamical modeling results (see Sect.\ref{bolom_mod}); 
slightly better fits are possible for later dates, but only
if the value of $\alpha$ approaches non-physical values. The full set
of model parameters which yields the minimum reduced $\chi^{2}$ is
given in Table~\ref{t:fits}, and the model curves are plotted in
Fig.~\ref{f:rlc}. For comparison, we show in Table~\ref{t:fits} the
equivalent parameters for three other well-sampled Type~IIb RSNe that were fitted using the
parametrization we used here:
SN~2001ig \citep{ryd04} in NGC~74724, SN~1993J \citep{kw07} in M81,
and SN~2001gd \citep{sto03} in NGC~5033. Fixing the value of $\delta$
to be $(\alpha - \beta -3)$, as in the \citet{che82} model for
expansion into a CSM with density decreasing as $r^{-2}$, also
resulted in a slightly better fit overall but a much steeper rise to
maximum that misses the earliest data at 5.5 and 9.0~GHz.

\begin{figure}
\includegraphics[width=8.cm]{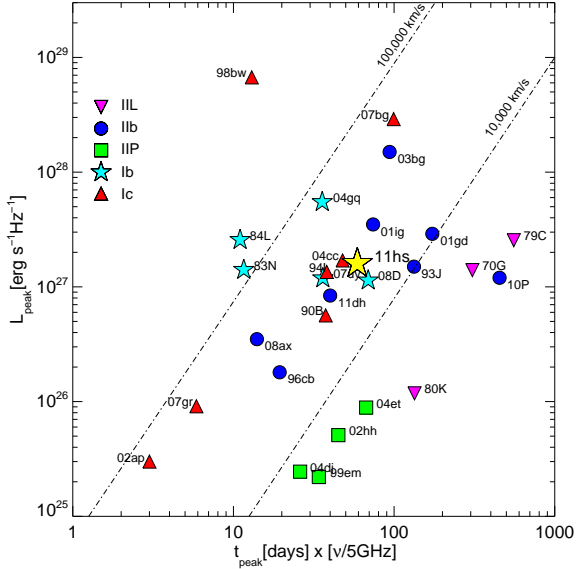}
  \caption{\label{radio_plot} Peak spectral radio luminosity at 5 GHz plotted versus the time of the peak at the same frequency measured for supernovae of Types Ib/c, IIb and IIL/P. The observed supernovae are designated by the last two digits of the year and letters. The references for the peak values used in this plot are in \citet{Romero}, where the radio observations towards SN\,2010P are also presented. Point-dashed lines show the mean velocity of the radio shell, with the assumption of the SSA as responsible for the flux peak (for details see \citealt{Chevalier98}).}
\end{figure}

\noindent
Both the optically-thin spectral index $\alpha$, and the rate of
decline $\beta$ are much steeper in SN~2011hs than in any of the other
Type IIb SNe, and the time to reach peak flux at  5~GHz is also much
shorter. \\
Using the methodology outlined in \citet{kw02} and \citet{sw03}, we
can derive an estimate of the progenitor's mass-loss rate, based on
its radio absorption properties. Substituting our model fit results
above into their equations~11 and 13 we find that

\begin{displaymath}
\frac{\rm \dot M/(M_{\odot}\ {\rm yr}^{-1})}{w /{\rm (10\,km\ s^{-1})}}
= (2.0 \pm 0.6) \times 10^{-5}
\end{displaymath}

\noindent
where $w$ is the mass-loss wind velocity, and the ejecta velocity
measured from the optical spectra (Fig.\,\ref{spec}) is in the range
$8,000-12,000$\,km\,s$^{-1}$. The deceleration parameter $m$ is given
by $m = -\delta/3$ (equation~6 of \citealt{kw02}) leading to $R
\propto t^{0.44}$ which is consistent with the rapid deceleration in
expansion velocity seen in Fig.~\,\ref{velox}. Despite similar light curve fitting parameters, the derived mass-loss rates for the
progenitors of SN 2011hs and SN 1993J are rather different, while those of SN 2001ig, and SN~2001gd are all remarkably
similar (see Table. \ref{t:fits}). 
In many respects SN~2011hs has behaved more like a Type Ib/c
SN than most ``normal'' Type~II SNe. The peak  luminosity at 5~GHz was
similar to that attained by SN~1993J, but only half as much as
SN~2001ig or SN~2001gd.
Recently, assuming the synchrotron self absorption (SSA) as the dominant absorption mechanism,  \citet{Chevalier10}
 estimated the radio shell velocity at the time of the peak radio luminosity (as in \citealt{Chevalier98}) for a sample
of SE SNe, and found that some of the Type IIb SNe show rapid radio evolutions for a given luminosity,  indicating high shell velocities (e.g.  around 30,000$-$50,000 km~s$^{-1}$) similar to Type Ib/c SNe.
Based on this, they propose dividing Type IIb SNe in the (L$_{peak}$ vs $t_{peak}$) plane (see their Fig.1, here adapted in Fig.\,\ref{radio_plot}) into two groups, eIIb and cIIb SNe, with Type eIIb SNe (like SNe 1993J and 2001gd) having more extended progenitor, a denser wind and a slower shock velocity, than those of Type cIIb (e.g. SNe 2008ax, 2003bg and 2001ig),  supposed to come from a more compact progenitor star.\\
Nevertheless, SN 2011dh was found to have {\it compact} radio properties \citep{Soderberg12}, while, as already discussed, there are strong evidences for an extended progenitor star (YSG; \citealt{Bersten}; \citealt{Ergon11dh}; \citealt{VanDyk11dhprog}). Note that the position of SN \,2001gd  changes if, instead of  \citet{sto03}, we adopt the results from \citet{sto07},  where a contribution from the SSA mechanism  in the modeling of the data was included. This assumption moves the position of SN 2001gd close to SN 2001ig, in the eIIb SNe ``region", making the separation even less definitive.
Similarly, SN\,2011hs seems to do not follow the proposed Type eIIb/cIIb separation: indeed, its position in the  (L$_{peak}$ vs $t_{peak}$) plot
is close to that of SN 2011dh in the {\it compact}  velocity contour, while the modeling  of the optical emission here presented, points to an {\it extended} progenitor.\\
 These findings seem to suggest that the radio emission is not a good indicator of the progenitor size,
although a larger sample of  Type IIb SNe observed at these wavelengths is needed to reach a firmer conclusion.\\ 


\section{Conclusions}\label{summary}
We have presented detailed spectrophotometric observations of SN\,2011hs taken from a few days after the explosion up to the nebular phase. 
The reported follow-up collects data from the X-ray to  radio wavelengths, turning this object into one of the most comprehensively studied Type IIb SNe.\\
We found that SN\,2011hs  was a relatively faint  (M$_B$ = -15.6 mag) and red Type IIb SN,  characterized  by a narrow light curve
shape, indicating a rapid evolution.
Its spectral evolution showed the metamorphosis typical of this class of SN, from spectra dominated by H\,I lines to
spectra where He\,I features dominate. The spectra are characterized by relatively broad absorption profiles from
which we measured high expansion velocities, similar to those of the fast expanding SN\,2003bg. 
This points to a high explosion energy per unit mass,
although the narrowness of the light curve and the faintness of the peak luminosity exclude 
the possibility of  a hypernova explosion.
The light curve shape could suggest a low mass progenitor
star, or more specifically a low ejecta density.  This could also explain the rapid evolution observed for the H\,I expansion velocity,
with the outer region in which the H lines form, receding faster than in previously studied SNe, probably because of a lower density. \\
Modelling the light curve of SN\,2011hs   and its velocity evolution with hydrodynamical calculations, we estimated that the
SN  is consistent with the explosion of a 3--4 M$_\odot$ He-core star, from a main sequence mass of 12--15 M$_\odot$, 
ejecting a $^{56}$Ni mass equal to 0.04 M$_\odot$  and characterized by an explosion energy  of $E\approx8.5\times10 ^{50}$ erg.
Such a scenario is also fully consistent with the results found by modeling the nebular spectrum taken at $\sim$215 days from maximum.
Based on different considerations on the light curve evolution, we assumed that the explosion epoch occurred 6 days before the discovery
(2,455,872 $\pm$ 4 JD). Such an explosion epoch, supported by the modelling of the radio light curve (from which we found an explosion occurred at $\ga$5 days
 before the discovery),  assumes an adiabatic cooling phase lasting 8 days, similar to that of SN\,1993J. Since the duration and the decreasing rate of the cooling branch depends mainly on the progenitor size,
we could infer from it a progenitor radius of $\approx$500--600 $R_\odot$. \\
We also analyze the possibility of a short rise time (with a 4-day long cooling phase),
giving an explosion scenario with the same He core mass (He3.3), slightly lower energy ($E= 6 \times 10^{50}$ erg) and the same $^{56}$Ni mass ejected ($\sim$0.04 $M_\odot$).
In contrast to the longer rise time models, we needed an extreme mixing (98\%) and a smaller radius ($\approx$ 300 $R_\odot$ ).
Although this case indicates the importance of an accurate estimation of the explosion time, the results
point again to a supergiant progenitor for SN\,2011hs.
Finally,  our modelling  rules out pre-explosion stars with He core mass $>5 M_\odot$, which implies excluding  main sequence masses above
$20$ $M_\odot$.  Such a lower limit for the progenitor mass 
 could indicate the possibility of a binary origin, although  the radio light curve
does not show strong deviations [as previously observed in e.g. SN\,2001ig \citep{ryd04} or SN 2003bg \citep{sod06}] 
as a signature of the presence of a companion star.
In summary, despite the fact that uncertainties in both the observational data (extinction, t$_{\mathrm{exp}}$, bolometric corrections for the early amateurs' points) and the modelling
prevent us from reaching definitive conclusions,  it stands out clearly that the SN\,2011hs progenitor  was a supergiant star 
with  ZAMS mass $<$20 M$_\odot$, as  found for SN 1993J, and most recently, for SN 2011dh.

\section*{Acknowledgments}
F.B. acknowledges support from FONDECYT through post-doctoral grant 3120227. F.B., G.P., S.G.G., J.P.A. and M.H. thank the support by the Millennium
Center for Supernova Science (P10-064-F), with input from ÔFondo de Innovaci\'on para la Competitividad,
del Ministerio de Economa, Fomento y Turismo de ChileÕ.
The authors acknowledge the Backyard Observatory Supernova Search (BOSS) team for their passionate effort and work in SN search, in particular Peter Marples (Loganholme Observatory, Queensland Australia), Greg Bock and Colin Drescher (Windaroo Observatory, Queensland, Australia).
 F.B. thanks  the Kavli Institute or the Physics and Mathematics of the Universe (Tokyo) for the hospitality and support during her visit while this paper was in progress. This research has been supported in part by WPI Initiative, MEXT, Japan.
E.C., M.T., S.B., P.M., are partially supported by the PRIN-INAF 2011 with the project ÓTransient Universe: from ESO Large to PESSTOÓ
S.G.G., J.P.A. and F.F. acknowledge support by CONICYT  through FONDECYT post-doctoral grant 3130680, grant 3110142 and grant 3110042, respectively. G.P. acknowledges partial support by ``Proyecto interno UNAB'' DI-303-13/R.
L.M. acknowledges financial support from Padua University grant CPS0204. LM acknowledges the Universidad Andr\'es Bello
in Santiago del Chile for hospitality while this paper was in progress.
E.P. is partially supported by grants  INAF PRIN 2009 and 2011 and ASI-INAF  I/088/06/0.
M.S. and C.C. gratefully acknowledge generous support provided by the Danish Agency for Science and Technology and Innovation  realized through a Sapere Aude Level 2 grant. C.R.C. is supported by the ALMA-CONICYT FUND Project 31100004.
M.H. ackowledges support from the John Simon Guggenheim Memorial Foundation. 
We are grateful to Elizabeth Mahoney, the ATCA staff, and numerous
numerous volunteer Duty Astronomers for assistance with carrying out
the radio observations remotely, often at short notice.  We thank Kurt
Weiler for providing the radio light curve fitting code.
This work is partially based on observations collected at the European Organization for Astronomical Research in the Southern Hemisphere, Chile (ESO), under the programs:
at NTT, ID 184.D-115 (P.I. S. Benetti); at VLT 089.D-0032 (P.I. P. Mazzali) and 090.D-0081 (P.I. E. Cappellaro). Observations were taken with REM, La Silla, Chile under the program AOT 24003 (P.I. F.Bufano). 
This paper is based on observations obtained through the CNTAC proposals CN2011B-092,  CN2011B-068, CN2012A-059.
Part of the optical/NIR photometry and spectroscopy presented in this paper were obtained by the Carnegie Supernova Project, which is supported by the National Science Foundation under Grant No. AST-1008343.


\label{lastpage}
\newpage

\begin{table*}
 \centering
 \begin{minipage}{140mm}
  \caption{ UVOT UV Observed Magnitudes of SN\,2011hs.\label{UV_mag}}
    \begin{tabular}{@{}lcccccc@{}}
     \hline
  UT & JD &     {\it w1}   & {\it w2}&   {\it m2} \\
  yyyy/mm/dd  &+2,400,000& mag& mag & mag& \\
  \hline
2011 Nov  15 &  55880.77 & 18.24  (0.10) & 19.07  (0.15) & 19.32  (0.26)   \\
2011 Nov 15 &  55881.17 & 18.18  (0.09) & 19.09  (0.14) & $\cdots$            \\
2011 Nov 18 &  55883.33 & 18.33  (0.10) & 19.01  (0.14) & 19.31  (0.19)  \\ 
2011 Nov 20 &  55885.75 & 18.55  (0.12) & 19.29  (0.17) & 19.80  (0.29)  \\ 
2011 Nov 22 &  55887.59 & 18.68  (0.15) & 19.44  (0.25) & 19.51  (0.30)  \\ 
2011 Nov 23 &  55889.39 & 18.93  (0.15) & 19.49  (0.19) & 19.71  (0.29)  \\ 
2011 Nov 26 &  55891.60 & 19.08  (0.17) & 19.50  (0.19) & 19.72  (0.27)  \\ 
    \hline
 \hline
\end{tabular}
\end{minipage}
\end{table*}



 \begin{table*}
 \centering
 \begin{minipage}{140mm}
  \caption{{\it UBVRI} photometric sequence in SN\,2011hs field as in Fig. \ref{FC}.\label{localJohn}}
    \begin{tabular}{@{}lccccccc@{}}
     \hline
  Star ID & R.A. & Dec. & {\it U} & {\it B} & {\it V} & {\it R} & {\it I}\\
               &h:m:s & $^\circ$:$'$:$''$& mag & mag & mag & mag & mag\\
               \hline
 1 & 22:57:21.390 & -43:21:59.51&  15.460  (0.022)  &   15.039(0.012)  & 14.188(0.025)  &    13.715(0.030)  &     13.290(0.025)\\ 
 2 & 22:57:22.294 & -43:23:16.01 &  $\cdots$ &  15.928(0.018)   &14.549(0.024)  &    13.677(0.034)  &     12.916(0.016)\\ 
 3 & 22:57:25.878 & -43:21:14.95 &  $\cdots$ &   18.449(0.032)   &17.935(0.033)  &    17.592(0.018)  &     17.216(0.059)\\ 
 4 & 22:57:22.824 & -43:26:23.77 &  $\cdots$ &   17.236(0.029)  & 16.804(0.020)  &    16.532(0.032)  &     16.229(0.020)\\ 
 5 & 22:57:23.152 & -43:27:11.07 &  $\cdots$ &   17.289(0.062)  &16.521(0.030)  &    16.067(0.036)  &     15.637(0.019)\\ 
 6 & 22:57:06.581 & -43:27:18.94 &  $\cdots$ &   15.832(0.026)   &14.985(0.030)  &    14.518(0.037)  &     14.119(0.019)\\
 7 & 22:56:57.117 & -43:27:41.50 &  $\cdots$ &   16.047(0.028)  & 15.038(0.036)  &   14.431(0.032)  &     13.910(0.016)\\  
 8 & 22:57:05.222 & -43:26:03.28 &  $\cdots$ &   19.514(0.062)   &18.338(0.061)  &    17.583(0.040)  &     16.939(0.031)\\ 
 9 & 22:56:58.012  &-43:26:04.85  & $\cdots$ &   18.738(0.020)   &18.119(0.040)  &    17.732(0.049)  &     17.326(0.054)\\
10 & 22:56:48.801 & -43:26:06.01 & $\cdots$ &   17.657(0.081)  & 17.077(0.033)  &    16.711(0.049)  &     16.357(0.030)\\
11 & 22:56:50.025 & -43:24:37.72 &  $\cdots$ &  17.930(0.066)   &17.198(0.034)  &    16.797(0.036)  &     16.386(0.027)\\ 
12 & 22:56:57.970  &-43:24:18.20 &  $\cdots$ &   17.941(0.035)  & 17.468(0.042)  &    17.156(0.039)  &     16.837(0.023)\\
13 & 22:56:58.428 & -43:22:57.74 & $\cdots$ &   17.868(0.071)   &16.753(0.025)  &    16.076(0.043)  &     15.522(0.026)\\ 
14 & 22:56:50.207&  -43:21:21.37 & $\cdots$ &   19.200(0.087)  & 18.049(0.058)  &    17.391(0.040)  &     16.694(0.020)\\ 
15 & 22:57:34.719 & -43:19:12.85 & $\cdots$ &   15.953(0.055)  & 15.415(0.023)  &    15.078(0.028)  &     14.729(0.026)\\ 
   \hline
 \hline
\end{tabular}
\end{minipage}
\end{table*}


 \begin{table*}
 \centering
 \begin{minipage}{140mm}
  \caption{{\it u$^\prime$g$^\prime$r$^\prime$i$^\prime$z$^\prime$} photometric sequence in SN\,2011hs field as in Fig. \ref{FC}.\label{localSloan}}
    \begin{tabular}{@{}lccccccc@{}}
     \hline
  Star ID & R.A. & Dec. & {\it u}$^\prime$ & {\it g}$^\prime$ & {\it r}$^\prime$ & {\it i}$^\prime$ & z$^\prime$\\
               &h:m:s & $^\circ$:$'$:$''$& mag & mag & mag & mag & mag\\
  \hline
 1 & 22:57:21.390 & -43:21:59.51&  16.359  (0.016)   & 14.570  (0.017)   & 13.951  (0.029)  &  13.731  (0.025)  &  13.613  (0.045) \\ 
 2 & 22:57:22.294 & -43:23:16.01 & 18.063  (0.042)   & 15.253  (0.014)   & 13.989  (0.032)  &  13.410  (0.025)  &  13.119  (0.034) \\ 
 3 & 22:57:25.878 & -43:21:14.95 &  $\cdots$             & 18.121  (0.032)   & 17.789  (0.021)  &  17.641  (0.033)  &  17.582  (0.079) \\ 
 4 & 22:57:22.824 & -43:26:23.77 & 17.779  (0.029)   & 16.963  (0.012)   & 16.712  (0.035)  &  16.633  (0.022)  &  16.574  (0.048) \\ 
 5 & 22:57:23.152 & -43:27:11.07 & 18.380  (0.104)   & 16.853  (0.021)   & 16.275  (0.026)  &  16.068  (0.022)  &  15.949  (0.030) \\ 
 6 & 22:57:06.581 & -43:27:18.94 & 17.264  (0.016)   & 15.355  (0.018)   & 14.733  (0.028)  &  14.555  (0.024)  &  14.477  (0.031) \\
 7 & 22:56:57.117 & -43:27:41.50 & 17.760  (0.019)   & 15.505  (0.010)   & 14.659  (0.031)  &  14.355  (0.023)  &  14.221  (0.046) \\  
 8 & 22:57:05.222 & -43:26:03.28 & 16.351  (0.010)   & 18.903  (0.020)   & 17.864  (0.042)  &  17.408  (0.065)  &  17.145  (0.015) \\ 
 9 & 22:56:58.012  &-43:26:04.85 & 19.817  (0.224)   & 18.352  (0.018)   & 17.946  (0.037)  &  17.757  (0.044)  &  17.678  (0.037) \\
10 & 22:56:48.801 & -43:26:06.01 & 18.332  (0.022)   & 17.303  (0.014)   & 16.905  (0.036)  &  16.787  (0.027)  &  16.714  (0.053) \\
11 & 22:56:50.025 & -43:24:37.72 & 18.866  (0.034)   & 17.495  (0.018)   & 16.994  (0.034)  &  16.811  (0.035)  &  16.720  (0.056) \\ 
12 & 22:56:57.970  &-43:24:18.20 & 18.465  (0.025)   & 17.671  (0.015)   & 17.364  (0.031)  &  17.245  (0.042)  &  17.190  (0.026) \\
13 & 22:56:58.428 & -43:22:57.74 & 19.932  (0.017)   & 17.280  (0.011)   & 16.334  (0.035)  &  15.981  (0.035)  &  15.798  (0.043) \\ 
14 & 22:57:34.719 & -43:19:12.85 &  $\cdots$  &  $\cdots$  & $\cdots$ & $\cdots$ & $\cdots$ \\ 
15 & 22:56:50.207&  -43:21:21.37 & 16.664  (0.018)   & 15.597  (0.013)   & 15.227  (0.029)  &  15.095  (0.02)7  &  15.088  (0.081) \\ 
   \hline
 \hline
\end{tabular}
\end{minipage}
\end{table*}


\begin{table*}
 \centering
 \begin{minipage}{140mm}
  \caption{{\it UBVRI} Observed Magnitudes of SN\,2011hs.\label{optical_mag}}
    \begin{tabular}{@{}lccccccc@{}}
     \hline
  UT & JD &  {\it U} & {\it B} & {\it V} & {\it R} & {\it I} & Instrument\\
  yyyy/mm/dd  &+2,400,000&mag & mag& mag&mag & mag& \\
  \hline
2011 Nov 12 &   55878.00   &    $\cdots$     &      $\cdots$   &           $\cdots$      &     15.53    (0.08)    &  $\cdots$    & BOSS    \\
2011 Nov 14 &   55879.90   &    $\cdots$     &      $\cdots$   &           $\cdots$      &     16.40    (0.07)    &  $\cdots$   & BOSS       \\
2011 Nov 14 &   55879.92   &    $\cdots$     &      $\cdots$   &           $\cdots$      &     16.42    (0.06)    &  $\cdots$   & BOSS       \\
2011 Nov 15 &   55880.57   &    $\cdots$     &  17.44    (0.02)  & 16.78    (0.01) &  16.40   (0.01) &  16.12     (0.01)& PROMPT \\
2011 Nov 15 &   55880.77   &  17.01 (0.07)& 17.40   (0.07) &  16.85    (0.08) &     $\cdots$          &     $\cdots$      & UVOT  \\
2011 Nov 15 &   55881.17   &  16.98 (0.07)& 17.29   (0.06) &  16.63    (0.07)&      $\cdots$          &     $\cdots$       & UVOT \\
2011 Nov 16 &   55881.64    &   $\cdots$     &  17.27    (0.02)  & 16.55    (0.01) &  16.14   (0.01)  & 15.93     (0.01) & PROMPT \\
2011 Nov 17 &   55882.53  &     $\cdots$     &  17.12    (0.02) &  16.34    (0.01) &  15.96    (0.01) &  15.73    (0.01) & PROMPT \\
2011 Nov 17 &   55882.58   &  17.02 (0.03)& 17.01   (0.02) &  16.39    (0.02) &  15.95    (0.02) &  15.72    (0.02)& EFOSC2\\ 
2011 Nov 18 &   55883.33  &   17.15 (0.08)& 16.94   (0.06) &  16.25    (0.06)&      $\cdots$           &   $\cdots$       & UVOT  \\
2011 Nov 18 &   55883.53&    $\cdots$    &   16.98    (0.02) &  16.18    (0.01)&   15.76    (0.01) &  15.54    (0.01)& PROMPT \\ 
2011 Nov 18 &   55883.54&    $\cdots$    &   16.97    (0.01) &  16.18    (0.01)&      $\cdots$           &   $\cdots$      &  CATA500   \\
2011 Nov 19 &   55884.57&    $\cdots$    &   16.96    (0.05)  & 16.02    (0.03) &  15.58    (0.01) &  15.37    (0.01)& PROMPT \\ 
2011 Nov 20 &   55885.53&    $\cdots$    &   16.97    (0.01) &  15.92    (0.01) &  15.48    (0.01) &  15.23    (0.01)& PROMPT  \\
2011 Nov 20 &   55885.75& 17.34 (0.08)&  16.86   (0.06) &  15.91   (0.06)&      $\cdots$          &    $\cdots$    & UVOT     \\
2011 Nov 22 &   55887.54&    $\cdots$    &   17.02    (0.01) &  15.80    (0.01) &     $\cdots$           &   $\cdots$       & PROMPT   \\
2011 Nov 22 &   55887.59& 17.67 (0.14)&  17.03   (0.08) &  15.83    (0.07)&      $\cdots$          &    $\cdots$      & UVOT   \\
2011 Nov 23 &   55889.35& 18.12 (0.13)&  17.19   (0.06) &  $\cdots$ &     $\cdots$          &    $\cdots$      & UVOT   \\
2011 Nov 24 &   55889.53&    $\cdots$    &      $\cdots$       &     15.80    (0.01) &  15.26    (0.01)  &    $\cdots$    & PROMPT     \\
2011 Nov 26 &   55891.53&    $\cdots$     &     $\cdots$       &     15.88    (0.01)  & 15.30    (0.01)  &    $\cdots$      & PROMPT    \\
2011 Nov 26 &   55891.60& 18.51 (0.16)&  17.37   (0.07)&   $\cdots$ &     $\cdots$          &    $\cdots$     & UVOT    \\
2011 Nov 28 &   55893.53&    $\cdots$    &   17.54    (0.02) &  16.02    (0.01) &  15.39    (0.01) &  15.01    (0.01)& PROMPT \\ 
2011 Nov 30 &   55895.58&    $\cdots$    &   17.80    (0.01) &  16.16    (0.01) &     $\cdots$           &   $\cdots$      &  CATA500   \\
2011 Dec 03 &   55898.58&    $\cdots$    &   18.19    (0.02) &  16.52    (0.01) &     $\cdots$           &   $\cdots$     &  CATA500    \\
2011 Dec 05 &   55900.68&    $\cdots$    &      $\cdots$         &   16.79    (0.01) &  15.99    (0.01)  & 15.55    (0.01) & PROMPT\\
2011 Dec 06 &   55901.53&    $\cdots$    &   18.44    (0.02) &  16.84    (0.01) &     $\cdots$           &   $\cdots$     &  CATA500   \\ 
2011 Dec 08 &   55903.54&    $\cdots$    &   18.63    (0.03)&   17.03    (0.01) &     $\cdots$           &   $\cdots$     &  CATA500    \\
2011 Dec 08 &   55903.58&    $\cdots$    &      $\cdots$          &  17.05    (0.01)  & 16.27    (0.01)  &    $\cdots$    & PROMPT    \\ 
2011 Dec 09 &   55904.54 &   $\cdots$    &   18.75    (0.04) &  17.14    (0.01) &     $\cdots$           &   $\cdots$     &  CATA500    \\
2011 Dec 10 &   55905.54&    $\cdots$    &   18.82    (0.04) &  17.17    (0.02)  &    $\cdots$           &   $\cdots$    &  CATA500     \\
2011 Dec 11 &   55906.62&    $\cdots$    &      $\cdots$           & 17.23    (0.03)  &    $\cdots$            &  $\cdots$        &  CATA500 \\
2011 Dec 13 &   55908.54&    $\cdots$    &   18.90    (0.04) &  17.34    (0.02) &     $\cdots$           &   $\cdots$       &  CATA500  \\
2011 Dec 15 &   55910.54&    $\cdots$    &   19.01    (0.03) &  17.39    (0.01) &     $\cdots$           &   $\cdots$      &  CATA500   \\
2011 Dec 15 &   55910.60&    $\cdots$    &   18.88    (0.04) &  17.45    (0.02) &  16.67   (0.02) &  16.07    (0.01) & PROMPT\\
2011 Dec 16 &   55911.56&    $\cdots$    &   18.83    (0.04) &  17.48  (0.02) &     $\cdots$            &  $\cdots$       &  CATA500  \\
2011 Dec 19 &   55914.56&  20.16 (0.06)&  19.03  (0.06) &  17.64    (0.06) &  16.84    (0.09) &  16.34    (0.10)& EFOSC2\\
2011 Dec 21 &   55916.54&  20.07 (0.05) & 19.06    (0.02) &  17.70    (0.02) &  16.88    (0.05) &  16.41    (0.03)& EFOSC2\\
2011 Dec 21 &   55916.54&    $\cdots$   &    19.03    (0.05)&   17.63    (0.05)   &   $\cdots$            &  $\cdots$     &  CATA500    \\
2011 Dec 26 &   55921.55&    $\cdots$   &    19.14    (0.06) &  17.77    (0.02)   &   $\cdots$            &  $\cdots$     &  CATA500    \\
2012 Apr 30 &   56047.89&    $\cdots$   &     20.93   (0.07)&   20.36    (0.11) &  19.67    (0.08)  & 19.22       (0.12)& WFCCD  \\
2012 May 02 &   56049.89&    $\cdots$   &     21.04   (0.11) &  20.41    (0.06)&   19.77    (0.07) &   $\cdots$   & WFCCD        \\
2012 Jun 20 &   56099.90&    $\cdots$    &    21.98   (0.09)&   21.56    (0.13)&   20.77    (0.06) &  20.33    (0.05)& FORS2  \\  
   \hline
 \hline
\end{tabular}
\end{minipage}
\end{table*}


\begin{table*}
 \begin{center}
 \begin{minipage}{140mm}
  \caption{{\it u$^\prime$g$^\prime$r$^\prime$i$^\prime$z$^\prime$} Observed Magnitudes of SN\,2011hs. \label{sloan_mag}}
    \begin{tabular}{@{}lccccccc@{}}
     \hline
  UT & JD &  {\it u}$^\prime$ & {\it g}$^\prime$ & {\it r}$^\prime$ & {\it i}$^\prime$& {\it z}$^\prime$ & Instrument\\
  yyyy/mm/dd  &+2,400,000&mag & mag& mag&mag & mag& \\
  \hline
2011 Nov 15 &   55880.56&  $\cdots$&  17.11    (0.02)  & 16.58    (0.02)  & 16.54    (0.02)&   16.50    (0.02)& PROMPT \\
2011 Nov 16 &   55881.66&  $\cdots$ & 16.90    (0.02) &  16.35    (0.02) &  16.31    (0.02) &  16.30    (0.02)& PROMPT \\
2011 Nov 17&    55882.55&  $\cdots$ & 16.70    (0.02)  & 16.14    (0.01) &  16.12    (0.02)&   16.04    (0.02)& PROMPT\\
2011 Nov 17 &   55883.56&  17.78 (0.04) &  16.53    (0.01) &  15.95    (0.01) &  15.92    (0.02) &  $\cdots$& CATA500 \\
2011 Nov 18&    55883.59&  $\cdots$ & $\cdots$& 15.95    (0.02) &  15.92    (0.02)  & 15.87    (0.02)& PROMPT \\
2011 Nov 19 &   55884.64&  $\cdots$ & $\cdots$& 15.79    (0.01) &  15.79    (0.02) &  15.70    (0.02)& PROMPT \\
2011 Nov 20 &   55885.54&  $\cdots$ & 16.41    (0.01) &  15.67    (0.01)&   15.63    (0.02)&   $\cdots$ &PROMPT \\
2011 Nov 21 &   55886.53&  $\cdots$ & 16.35    (0.01) &  $\cdots$& $\cdots$& $\cdots$ &PROMPT \\
2011 Nov 22  &  55887.54&  $\cdots$ & 16.35    (0.02) &  $\cdots$& $\cdots$ & $\cdots$ &PROMPT \\
2011 Nov 23 &   55888.59&  $\cdots$&  $\cdots$ &15.49    (0.01) &  15.40    (0.01)&   15.33    (0.02) &PROMPT \\
2011 Nov 24 &   55889.53&  $\cdots$&  16.42    (0.01) &  $\cdots$ &  $\cdots$& $\cdots$ & PROMPT \\
2011 Nov 25 &   55890.54&  $\cdots$&  16.48    (0.02) &  15.47    (0.01) &  $\cdots$ &  $\cdots$ & PROMPT \\
2011 Nov 26 &   55891.54&  $\cdots$&  16.54    (0.01) &  $\cdots$ &  $\cdots$& $\cdots$ & PROMPT \\
2011 Nov 27 &   55892.56&  $\cdots$&  $\cdots$& $\cdots$ &15.42    (0.03) &  15.36    (0.02) & PROMPT\\
2011 Nov 28 &   55893.54&  $\cdots$&  16.65    (0.02) &  $\cdots$& $\cdots$& $\cdots$ &PROMPT \\
2011 Nov 29 &   55894.53 & $\cdots$&  16.77    (0.01)&   $\cdots$& $\cdots$& $\cdots$ &PROMPT \\
2011 Nov 29 &   55895.62 & 19.53 (0.14)&   16.93    (0.01) &  15.78    (0.01)  & 15.55    (0.01) &  $\cdots$ & CATA500 \\
2011 Dec 02 &   55897.54&  $\cdots$&  17.08    (0.01) &  $\cdots$& $\cdots$& $\cdots$ &PROMPT \\
2011 Dec 02 &   55898.58&  19.98 (0.14)&   17.29    (0.01)  & 16.07    (0.01)  & 15.77    (0.01) &  $\cdots$& CATA500 \\
2011 Dec 04&    55899.57&  $\cdots$&  $\cdots$& 16.18    (0.02) &  15.92    (0.02) &  15.75    (0.02)& PROMPT\\
2011 Dec 05&    55901.57&  $\cdots$&   17.64    (0.02) &  16.38    (0.02) &  16.05    (0.02)  & $\cdots$& CATA500\\
2011 Dec 07 &   55903.58&  20.16 (0.21) &  17.77    (0.02)  & 16.55    (0.01) &  16.16    (0.01)  & $\cdots$& CATA500 \\
2011 Dec 08 &   55904.58&  20.35 (0.26) &  17.83    (0.02) &  16.59    (0.01) &  16.24    (0.02)  & $\cdots$& CATA500\\
2011 Dec 09 &   55904.58&  $\cdots$&  $\cdots$ &  $\cdots$ &16.30    (0.02) &  16.04    (0.02)& PROMPT \\
2011 Dec 09 &   55905.58&  20.24 (0.27) &  17.92    (0.02) &  16.66    (0.01) &  16.30    (0.02) &  $\cdots$& CATA500\\
2011 Dec 10 &   55906.66&  $\cdots$&  18.05    (0.04)&   16.78    (0.02) &  16.39    (0.02) &  $\cdots$ &CATA500\\
2011 Dec 14 &   55909.61&  $\cdots$&  18.13    (0.07) &  $\cdots$& $\cdots$ &  $\cdots$ &PROMPT \\
2011 Dec 22 &   55918.55&  $\cdots$&  18.26    (0.04) &  17.22    (0.03) &  16.73    (0.08) &  $\cdots$ &CATA500 \\
2012 Jan 14 &   55940.57&  $\cdots$&  18.57    (0.09) &  17.81    (0.05) &  $\cdots$& $\cdots$ &PROMPT\\
2012 Jan 16 &   55942.56&  $\cdots$&  $\cdots$ &  17.91    (0.02) &  $\cdots$ &  $\cdots$ &PROMPT\\
  \hline
 \hline
\end{tabular}
\end{minipage}
\end{center}
\end{table*}


\begin{table*}
 \centering
 \begin{minipage}{140mm}
  \caption{NIR Observed Magnitudes of SN\,2011hs.\label{NIR_mag}}
    \begin{tabular}{@{}lccccccc@{}}
     \hline
  UT & JD &  {\it J} & {\it H} & {\it K} & Instrument\\
  yyyy/mm/dd  &+2,400,000& mag& mag& mag& \\
  \hline
2011 Nov 18  &  55883.57 &  15.14  (0.06) &  15.08  (0.04)   & 14.81  (0.05)  &   SOFI  \\
2011 Dec 07  &  55902.61 &  15.22  (0.04)  & 14.85  (0.09)    & $\cdots$ &   RetroCam \\
2011 Dec 11  &  55906.55 &  15.50  (0.17)  &  15.10  (0.18)   &  $\cdots$&   REM \\
2011 Dec 12  &  55907.58 &  15.42  (0.11)  & 14.99  (0.05)   &  $\cdots$&   RetroCam \\
2011 Dec 13  &  55908.58 &  15.46  (0.09)  & 15.16  (0.17)   &  $\cdots$ &   RetroCam \\
2011 Dec 15  &  55910.58 &  $\cdots$ & 15.18  (0.09)   &  $\cdots$ &   RetroCam \\
2011 Dec 20 &   55916.56 &  15.76  (0.05)  &  15.53  (0.14)  &  15.13  (0.06)  &   SOFI  \\
2011 Dec 25 &   55921.56 &  15.92  (0.22)  &   $\cdots$&   $\cdots$&   REM  \\
2011 Dec 28 &   55924.56 &  15.85  (0.13)  &   $\cdots$&   $\cdots$&   REM  \\
    \hline
 \hline
\end{tabular}
\end{minipage}
\end{table*}



 \begin{table*}
 \centering
 \begin{minipage}{140mm}
  \caption{Journal of Optical and NIR Spectroscopic Observations of SN\,2011hs.\label{spec_journal}}
  \begin{tabular}{@{}lcccc@{}}
   \hline
Date  & JD & Phase\footnote{Phase in days from B band maximum light ($t(B)_{max}$ = 2,455,885.5$\pm$ 1.0 JD).} & Instrumental & Spectral Range\\
yyyy/mm/dd & +2,400,000 & days & set up & \AA \\
  \hline
 2011 Nov 14   & 55880.4  &-5.1& SALT+RSS+p0900&3500-9000\\
 2011 Nov 15   & 55880.5  &-5.0& Magellan+LDSS3+VPH-All&3700-9000\\
 2011 Nov 15   & 55881.3  &-4.2&SALT+RSS+p0900&3900-8900\\
 2011 Nov 16   & 55881.6  &-3.9&SOAR+Goodman+300 l/mm&3900-8900\\ 
 2011 Nov 17   & 55882.5  &-3.0&NTT+EFOSC2+Gr\#13&3650-9250\\
 2011 Nov 18   & 55883.5  &-2.0&NTT+SOFI+BG+RG&9400-23000\\
 2011 Nov 18   & 55883.5  &-2.0&du Pont+WFCCD+GrismBlue&3650-9250\\
 2011 Nov 18   & 55883.5  &-2.0&Magellan+IMACS+300 l/mm&4200-9300\\
 2011 Nov 27   & 55892.5  &+7.0&du Pont+WFCCD+GrismBlue&3650-9200\\
 2011 Nov 28   & 55893.4  &+7.9&du Pont+WFCCD+GrismBlue&3650-9200\\
 2011 Dec 01   & 55897.3  &+11.7&SALT+RSS+p0900&3600-9000\\
 2011 Dec 18   & 55913.5  &+28.0&Magellan+FIRE+LongSlit&8050-25000\\
 2011 Dec 18   & 55913.5  &+28.0&SOAR+Goodman+300 l/mm&3900-8900\\
 2011 Dec 19   & 55914.5  &+29.0&NTT+EFOSC2+Gr\#13&3650-9250\\
 2011 Dec 20   & 55915.5  &+30.0&NTT+SOFI+BG+RG&9700-23000\\
 2011 Dec 21   & 55916.5  &+31.0&NTT+EFOSC2+Gr\#13&3650-9250\\
 2011 Dec 21   & 55916.5  &+31.0&Magellan+FIRE+LongSlit&8050-25000\\
 2012 Jan 15    & 55941.5  &+56.0&SOAR+Goodman+300 l/mm&4000-8900\\
 2012 Jan 20    & 55946.5  &+61.0&Magellan+LDSS3+VPH-All&3650-9400\\
 2012 May 01   & 56048.9  &+163.4&NTT+EFOSC2+Gr\#13&3650-9000\\
 2012 Jun 21  &56099.9&+214.4& VLT+FORS2+V300&3950-9600\\
 2012 Jun 23  &56101.8&+216.3&  Magellan+LDSS3+VPH-All&3850-9800\\
 2012 Oct 22  &56222.5&+337& VLT+FORS2+V300&3950-9600\\
    \hline
 \hline
\end{tabular}
\end{minipage}
\end{table*}


\begin{table*}
\centering
\begin{minipage}{140mm}
\caption{SN\, 2011hs radio flux measurements with the ATCA.\label{t:radiofluxes}}
\begin{tabular}{lcccccc}
\hline
Date     & \multicolumn{1}{c}{Days since} & $S$(18.0\, GHz) & $S$(9.0~GHz) &
 $S$(5.5~GHz) &    $S$(2.7~GHz) & $S$(2.0~GHz) \\
yyyy/mm/dd    & discovery & \multicolumn{1}{c}{mJy}     &
 \multicolumn{1}{c}{mJy} & \multicolumn{1}{c}{mJy}    &
 \multicolumn{1}{c}{mJy} & \multicolumn{1}{c}{mJy}    \\
\hline
2011 Nov 17 &      5   &      $\cdots$     &   $<0.15$      &   $<0.15$      &         $\cdots$     &     $\cdots$      \\
2011 Dec 24 &     12   &      $\cdots$     & $0.65\pm0.13$  & $0.17\pm0.05$  &         $\cdots$     &     $\cdots$      \\
2011 Dec 04 &     22   & $1.44\pm0.15$  & $1.60\pm0.10$  & $0.57\pm0.05$  &        $\cdots$     &     $\cdots$      \\
2011 Dec 14 &     32   & $1.32\pm0.13$  & $1.98\pm0.12$  & $1.69\pm0.10$  &        $\cdots$     &     $\cdots$      \\
2011 Dec 23 &     41   & $1.04\pm0.11$  & $1.92\pm0.11$  & $1.78\pm0.10$  &        $\cdots$     &     $\cdots$      \\
2012 Jan 03 &     52   & $0.28\pm0.04$  & $1.54\pm0.09$  & $1.88\pm0.10$  &        $\cdots$     &     $\cdots$      \\
2012 Jan 11 &     62   & $0.19\pm0.04$  & $1.26\pm0.08$  & $1.71\pm0.10$  &        $\cdots$     &     $\cdots$      \\
2012 Feb 03 &     85   & $\cdots$          & $1.18\pm0.07$  & $1.72\pm0.09$  & $1.54\pm0.16$     & $1.05\pm0.11$ \\
2012 Mar 04 &    115   & $\cdots$          & $0.51\pm0.06$  & $1.23\pm0.08$  & $1.11\pm0.12$     & $1.33\pm0.20$ \\
2012 May 05 &    177   & $\cdots$          & $0.20\pm0.03$  & $1.11\pm0.08$  & $1.62\pm0.17$     & $1.60\pm0.17$ \\
\hline
\end{tabular}
\end{minipage}
\end{table*}


\begin{table*}
 \begin{minipage}{85mm}
  \caption{Light-curve parameters of SN\,2011hs.\label{maximum}}
    \begin{tabular}{@{}ccccc@{}}
  Filter & $t(\lambda)_{max}$ & Maximum Mag. & Rise Rate\footnote{Decline within 5 days before $t(\lambda)_{max}$}& Decline Rate \footnote{Decline within 15 days after $t(\lambda)_{max}$.} \\
               & JD+2,400,000 & mag & mag/100d & mag/100d \\
               \hline
    U&55,881.5$\pm$1.0&17.00$\pm$0.02& $\cdots$&16.2$\pm$1.7\\  
    B &55,885.5$\pm$1.0& 16.93$\pm$0.01& $-9.2\pm$0.4   &9.0$\pm$0.1\\
    V &55,888.5$\pm$1.0& 15.80$\pm$0.01&  $-7.9\pm$0.2  &9.0$\pm$0.1\\
    R &55,889.5$\pm$1.0& 15.28$\pm$0.01&$-5.9\pm$0.3&7.4$\pm$0.1\\
    I  &55,890.5$\pm$1.0&15.01$\pm0.01$& $-5.4\pm$0.2   &7.4$\pm$0.2\\
    \hline
    {\it g$^\prime$}& 55,887.5$\pm$1.0 &16.35$\pm$0.01& $-6.8\pm$0.4  &9.2$\pm$0.1\\
   {\it r$^\prime$} &55,889.5$\pm$1.0&15.46$\pm$0.01& $-6.3\pm$0.3  &8.5$\pm$0.1\\   
   {\it i$^\prime$} &55,890.5$\pm$1.0&15.36$\pm$0.01& $-5.1\pm$0.4  &7.7$\pm$0.1\\  
  {\it z$^\prime$} &55,890.5$\pm$1.0&15.30$\pm$0.02& $\cdots$ &5.6$\pm$0.2\\  
    \hline
    \hline      
    
\end{tabular}
\end{minipage}
\end{table*}


\begin{table*}
\centering
\begin{minipage}{140mm}
\caption{Comparison of radio light curve model parameters.\label{t:fits}}
\begin{tabular}{ccccc}
\hline
Parameter    &     SN 2011hs     &      SN 2001ig    &      SN 1993J      &    SN 2001gd  \\
\hline
$K_1$        & $5.2\times10^{3}$ & $2.7\times10^{4}$ & $4.8\times10^{3}$ & $1.5\times10^{3}$ \\
$\alpha$     &     $-1.90$      &      $-1.06$      &     $-0.81$       &     $-1.38$        \\
$\beta$      &     $-1.66$      &      $-1.50$      &     $-0.73$       &     $-0.96$        \\
$K_2$        & $1.5\times10^{2}$ & $1.4\times10^{3}$ & $1.6\times10^{2}$  & $3.3\times10^{6}$ \\
$\delta$     &     $-1.31$      &     $-2.56$      &     $-1.88$       &      --            \\
$K_3$        & $1.9\times10^{5}$ & $1.5\times10^{5}$ & $4.3\times10^{5}$ & $1.1\times10^{3}$ \\
$\delta^{\prime}$ &  $-3.06$      &      $-2.69$     &     $-2.83$       &     $-1.27$        \\
$K_4$        &        0.0        &        0.0       &      0.0         &      --           \\
Time to $L_{\rm 5\ GHz \ peak}$ (days) & 59 &  74       &      133         &      173            \\
$L_{\rm 5\ GHz \ peak}$ (erg s$^{-1}$ Hz$^{-1}$) &
          $1.6\times10^{27}$     &   $3.5\times10^{27}$ & $1.5\times10^{27}$   & $2.9\times10^{27}$ \\
Mass-loss rate ($w/10$~km~s$^{-1}$) M$_{\odot}$ yr$^{-1}$ &
  $(2.0\pm0.6)\times10^{-5}$                &      $(2.2\pm0.5)\times10^{-5}$  & $0.5-5.9\times10^{-6}$ & $3.0\times10^{-5}$ \\
\hline
\end{tabular}
\end{minipage}
\end{table*}

\end{document}